\documentclass[sigconf]{acmart}
\AtBeginDocument{%
  }

\setcopyright{acmlicensed}
\copyrightyear{2026}
\acmYear{2026}
\acmDOI{XXXXXXX.XXXXXXX}
\acmConference[CCS '26] {Proceedings of the 2026 ACM SIGSAC Conference on Computer and Communications Security}{November 15--19, 2026}{The Hague, Netherlands.}
\acmISBN{978-1-4503-XXXX-X/2018/06}

\renewcommand\footnotetextcopyrightpermission[1]{} 
\usepackage{balance}
\usepackage{multirow, multicol}
\usepackage{graphicx}
\usepackage[normalem]{ulem}
\usepackage{subcaption}

\begin{document}

\title{Adaptive Diffusion Freezing: Privacy-preserving Diffusion Models Against Membership Inference Attacks}

\author{Jialu Guo}
\affiliation{%
  \institution{MIIT Key Laboratory of Data Intelligence and Management}
  \institution{Beihang University}
  \city{Beijing}
  \country{China}
}
\email{jialurita@gmail.com}


\author{Xiao Han}
\affiliation{%
  \institution{MIIT Key Laboratory of Data Intelligence and Management}
  \institution{Beihang University}
  \city{Beijing}
  \country{China}
}
\email{xh\_bh@buaa.edu.cn}

\author{Junjie Wu}
\affiliation{%
  \institution{MIIT Key Laboratory of Data Intelligence and Management}
  \institution{Beihang University}
  \city{Beijing}
  \country{China}
}
\email{wujj@buaa.edu.cn}

\renewcommand{\shortauthors}{Jialu Guo, Xiao Han, and Junjie Wu}

\begin{abstract}
    Diffusion models have achieved remarkable success in generative tasks across various areas, however their training process raises significant privacy concerns, particularly under membership inference attacks (MIAs). Prior studies on privacy-preserving of diffusion models fail to balance privacy, utility, and efficiency. To address this gap, we propose a novel framework of privacy-preserving diffusion models, Adaptive Diffusion Freezing (ADF), which can defend against MIAs with better trade-off. By leveraging cross-timestep adaptive freezing training, ADF explicitly control the participation of different data subsets across diffusion timesteps via a mask matrix, which reduces the over-memorization and leads to more uniform model behaviors between member and nonmember samples. To construct a freezing mask matrix that effectively reduce membership leakage without unnecessarily harming generation quality, we introduce a pretraining-based risk-aware freezing policy to estimate MIA risk based on memorization tendency, and suppress the contribution of the subset-timestep pairs with higher risk. Evaluations on multiple datasets demonstrate that ADF provides effective defense performance as well as state-of-the-art privacy-utility-efficiency trade-off performance compared to various baselines.
\end{abstract}



\maketitle

\section{Introduction}

The rapid advancement of generative models has significantly expanded their applicability across diverse domains. With powerful data modeling and synthetic capability, generative models provide an effective solution to the challenges of costly, sensitive, and inherently imbalanced data acquisition in real-world applications. Among generative models, diffusion models have emerged as a leading paradigm. Recent studies have discovered the limitations of traditional generative models. For examples, variational autoencoders, although easier to train, often suffer from limited generation quality due to blurry reconstructions\cite{bredell2023explicitly}; generative adversarial networks are well known to suffer from unstable training dynamics, such as mode collapse\cite{saxena2021generative,dhariwal2021diffusion}.
In contrast, diffusion models achieve high sample fidelity and diversity with a more stable training process~\cite{ho2020denoising,dhariwal2021diffusion,cao2024survey,yang2023diffusion}. These advantages have enabled their adoption in various domains, such as synthesizing rare disease samples, mitigating data scarcity in healthcare, and alleviating cold-start problems in market analysis~\cite{cao2024survey}. 

Despite these advantages, the deployment of diffusion models raises critical privacy and security concerns, including privacy violations and copyright infringement~\cite{manduchi2024challenges}. Recent privacy regulations and AI governance frameworks, such as GDPR, emphasize that machine learning models should not expose private information contained in their training datasets. In this context, membership inference attacks (MIAs) have posed a representative threat toward training-data privacy. Given access to a target model, MIA aims to determine whether a specific sample was included in the model's training set. A successful MIA may reveal the participation of sensitive individual records or proprietary data, thereby leading to serious privacy leakage. Recent studies have shown that diffusion models are also vulnerable to MIAs~\cite{sun2023attribute,duan2023diffusion,matsumoto2023membership,pang2023white}. These findings highlight the necessity of developing effective MIA defenses for diffusion models, especially when they are trained on sensitive data and deployed for data sharing. 

In addition, MIAs against diffusion models differ from conventional MIAs against discriminative models. Traditional MIAs typically exploit discrepancies in model outputs, confidence scores, or losses between member and nonmember samples. These discrepancies are usually caused by global overfitting, where the target model memorizes training samples and therefore behaves differently on members than on nonmembers. In contrast, diffusion models are trained through a sequence of timestep-dependent denoising tasks, where each training sample repeatedly contributes across the denoising trajectory. As a result, diffusion-specific MIAs rely on the behavioral discrepancy accumulated over timesteps, using timestep-dependent signals such as denoising error, generative likelihood, or gradients, to distinguish member samples from nonmember samples~\cite{duan2023diffusion,hu2023membership,pang2023white}. This difference indicates that membership leakage in diffusion models is not merely caused by global overfitting, but is closely coupled with the multi-timestep denoising architecture~\cite{duan2023diffusion,matsumoto2023membership}. Different timesteps may exhibit different levels of memorization tendency and may overfit different portions of the training data to different extents. Therefore, defending MIAs in diffusion models requires a fine-grained strategy that accounts for timestep-level over-memorization heterogeneity, rather than globally weakening the entire training process.

Unfortunately, existing MIA defenses for diffusion models are very limited. On the one hand, several privacy-preserving techniques that have been proposed to defend against MIAs in general machine learning models, including data augmentation~\cite{devries2017improved,duan2023diffusion}, regularization~\cite{pang2023white}, and differential privacy~\cite{abadi2016deep,pang2023white}, can be directly applied to diffusion models, but their effectiveness is often constrained. The key reason is that these defenses largely ignore a defining characteristic of diffusion models: multi-timestep denoising, which is critical to MIAs on diffusion models. More specifically, regularization-based defenses mitigate MIAs by penalizing overfitting, yet in diffusion models a sample may overfit severely at specific denoising timesteps while still appearing well-fitted on average across timesteps. Such samples can therefore evade conventional regularization while remaining susceptible to diffusion-specific MIAs. Similarly, existing differential privacy and data augmentation mechanisms are not designed to suppress timestep-wise membership leakage. This mismatch between the timestep-level of diffusion-specific MIAs and the model-level design of existing defenses motivates the need for a defense mechanism tailored to the denoising dynamics of diffusion models.


In addition, only a few studies specifically focus on defending against MIAs in diffusion models, such as knowledge distillation~\cite{fernandez2023privacy} and adversarial training~\cite{luo2025privacy}. However, these methods largely inherit general privacy-preserving strategies from conventional machine learning, and do not explicitly address the timestep-wise membership leakage signals unique to diffusion models. Specifically, they suppress membership leakage at the model level. For example, knowledge distillation transfers knowledge from a teacher model to a student model, and adversarial training tempts to improve the robustness against privacy attacks, rather than identifying how overfitting emerges across denoising timesteps and eliminating attack signals. 
Beyond this fundamental limitation, existing defenses also face practical constraints. For instance, knowledge distillation typically requires training additional teacher-student models, which introduces extra training complexity and computational overhead. In addition, existing adversarial training method is primarily designed for fine-tuning pre-trained latent diffusion models (LDMs), which protects only fine-tuning data while leaving the original training dataset vulnerable. These limitations highlight the need for a diffusion-specific defense that can mitigate timestep-wise membership leakage while balancing privacy preservation, generative utility, and computational efficiency.

This work aims to develop a privacy-preserving framework tailored to the unique training paradigm of diffusion models, aiming to address MIA risk by effectively regulating the over-memorization of training samples at timestep-level. However, realizing such intuition in diffusion models is non-trivial, and requires addressing the following key challenges.

\textit{Challenge 1: How to estimate the over-memorization at each timestep and training subsets?} As mentioned above, memorization in diffusion models is jointly dependent on both data and timesteps. Since each sample contributes across the whole denoising trajectory, over-memorization is coupled with the temporal structure of diffusion training rather than arising as a purely global phenomenon, making it insufficient to assess privacy risk only at the model-wide level. Therefore, an effective defense must be able to identify memorization at an appropriate granularity, namely across different training subsets and different timesteps.

\textit{Challenge 2: How to suppress over-memorization while maintaining generation quality?} By addressing the first challenge, we may estimate the memorization toward certain subset-timestep pairs in diffusion training. But, it is still unclear how to suppress it without harming generative utility. Considering the different overfitting level across timesteps, a straightforward approach would be to entirely remove those timesteps that exhibit higher over-memorization. However, this naive solution severely disrupts the temporal dependency of diffusion learning, since the denoising transition between adjacent timesteps is intrinsically connected. Once a timestep $t$ is entirely omitted, the model can no longer properly learn the corresponding denoising mapping from $\mathbf{x}_{t} \to \mathbf{x}_{t-1}$, which may significantly degrade generative utility. Therefore, a practical defense should only partially reduce training exposure at a timestep, especially for those data subsets that are more prone to memorization, so that privacy protection can be achieved without destroying the overall denoising dynamics. 

\textit{Challenge 3: How to realize subset-timestep control efficiently within standard diffusion training?} Even if high-risk subset-timestep pairs can be identified, implementing such control during diffusion training is non-trivial. DDPM training relies on randomly sampled timesteps and mini-batches, where samples with different memorization tendency are mixed in each update. Simple implementation may require costly per-sample risk estimation or complex reweighting strategies, leading to substantial computational overhead. Moreover, unstructured suppression may create imbalanced supervision across timesteps and destabilize denoising training. Therefore, an effective defense should realize subset-timestep level control in a lightweight and structured manner, while remaining compatible with the standard DDPM training objective.

To address these challenges, we propose a novel defense framework, Adaptive Diffusion Freezing (i.e., ADF). The central idea of ADF is to explicitly regulate over-memorization at the level of subset-timestep pairs during diffusion training. Instead of allowing all training samples to participate uniformly across all timesteps, ADF adaptive freezes the contribution of certain training subsets at certain timesteps, thereby preventing memorization-prone signals from exerting disproportionate influence on the model. Concretely, ADF is built upon two key design mechanisms. Firstly, we introduce the cross-timestep adaptive freezing training, which adaptively restricts the exposure of training samples across diffusion timesteps. Instead of allowing all samples to participate uniformly throughout the denoising trajectory, ADF uses a timestep-wise mask matrix to control whether a subset contributes to optimization at a given timestep, thereby preventing the model from over-memorizing specific instances. Second, this masking strategy is guided by a risk-aware freezing policy, which estimates the MIA risk of different data subsets based on the memorization tendency of different subset-timestep pairs and allocates the freezing budget accordingly, so that the defense can focus on more memorization-prone training signals without unnecessarily sacrificing generative quality. In addition, our framework can be extended to a multi-network parameterization scheme, where different timestep groups are modeled by separate networks, further enhancing flexibility and privacy control. Overall, ADF effectively mitigates membership leakage in diffusion models while preserving generative performance and computational efficiency.

The main contributions in this paper include: 

1) We propose a novel framework, Adaptive Diffusion Freezing (ADF) , for defending MIAs in diffusion models, and achieve the best trade-off between privacy, utility, and efficiency.

2) We offer a fine-grained and dynamically controllable defense mechanism through adaptive freezing training and risk-aware mask construction, enabling flexible adjustment of the privacy-utility-efficiency trade-off via sparsity control and timestep-aware network design. 

3) We conduct extensive experiments and implement the proposed method to train diffusion model 
on four datasets and compare our method with five baseline methods under various MIA attacks to verify the superiority of our method.

\section{Related Work}
In this section, we review the existing research related to the membership inference attack (MIA) defense for diffusion models. Firstly, we begin with an overview of diffusion models, with particular focus on DDPM, which serves as the primary diffusion backbone in our main experiments. Second, we summarize the various types of membership inference attacks on diffusion models. Finally, we examine prior defense mechanisms and analyze their limitations, which serve as the motivation for our proposed defense approach.

\subsection{Diffusion Models}
Diffusion models is a class of deep generative models that involves two main processes: forward diffusion and reverse denoising, to transform random noise into meaningful data~\cite{truong2025attacks}. The forward diffusion process starts from a data sample and progressively adds Gaussain noise in predefined $T$ timesteps. After transforming the original data into pure Gaussian noise, conversely, the reverse denoising process aims to undo the transformation by training a neural network to learn to gradually denoise the noisy sample. Generally, existing diffusion-based generative models can be broadly summarized into three categories, i.e., Denoising Diffusion Probabilistic Model (DDPM), Noise Conditioned Score Networks (NCSN), and Score Stochastic Differential Equation (SDE). DDPM is proposed by Ho and Jain~\cite{ho2020denoising}, and introduces a Markov chain to gradually add Gaussian noise to original sample and trains a neural network to learn the denoising distribution. DDPM achieves a high-quality, efficient, and controllable diffusion model framework for generative tasks, and serves as the foundation for many subsequent diffusion-based models~\cite{song2020denoising,song2020score}. Therefore, we mainly focus on the DDPM framework in this paper, designing and employing our defense method based on its standard training paradigm. Nevertheless, ADF is not restricted to DDPM, and we further validate its generalizability to other mainstream diffusion architectures.

\subsection{Membership Inference Attacks toward Diffusion Models}
Membership inference attack (MIA) is a privacy attack toward deep learning models, in which the adversary aims to determine whether a given sample was in the training dataset of the target model~\cite{shokri2017membership}. It was first proposed and studied on classification models and have been recently targeted on generative models~\cite{hu2022membership}, i.e., GANs~\cite{chen2020gan}, diffusion models~\cite{fu2025unlocking}. According to the preliminary knowledge of the adversary, the MIAs on diffusion models can be divided into three types: black-box, gray-box, and white-box~\cite{truong2025attacks}.

\textbf{Black-box attacks.} 
In black-box attack, an adversary can only query the target model and observe the generated outputs, without access to model parameters or internal states. Existing methods typically rely on reconstruction similarity or distributional discrepancies between generated samples and candidate inputs. For instance, Matsumoto et al.~\cite{matsumoto2023membership} extended the full black-box attack framework proposed by Chen et al.~\cite{chen2020gan} on GANs to DDIM, demonstrating that although such attacks are feasible, their effectiveness is significantly reduced due to the complex multi-step generation process of diffusion models. Concurrently, other works attempt to improve attack performance under black-box setting by leveraging auxiliary datasets or introducing additional assumptions~\cite{li2024towards,zhang2024generated,wu2022membership}, such as the access to partial training data or the substitution between synthetic dataset and training dataset, which restrict the practicality.

\textbf{Gray-box attacks.} 
The diffusion model generates images through multiple denoising timesteps and produces many intermediate outputs, s. Therefore, the gray-box attack assumes that attackers can obtain the added noise and intermediate outputs at each timestep to derive richer attack signals, but still have no prior knowledge, e.g., parameters, gradients. For example, Duan et al.~\cite{duan2023diffusion} proposed SecMI based on t-error, which is the matching loss of forward process posterior estimation at each timestep, 
assuming that t-error of member is smaller than that of nonmember.

\textbf{White-box attacks.}
In the white-box setting, attackers have full access to the target model, including the parameters, gradients, architecture, source code, etc, which enables more powerful attacks. Prior works exploited the observation that member samples typically yield lower training loss and higher generated likelihood due to the overfitting of the target model~\cite{matsumoto2023membership,hu2023loss}. Additionally, Pang et al.~\cite{pang2023white} introduced GSA, which further extracts high-dimensional gradient features to improve attack accuracy.

\subsection{Defense Mechanisms Against MIA}
To mitigate privacy risks caused by MIAs, recent studies have employed prior defense approaches or proposed new mechanisms for diffusion models~\cite{truong2025attacks}, including data augmentation, regularization, differential privacy, knowledge distillation, and adversarial training. 

Regularization and data augmentation reduce overfitting via constraints to model complexity~\cite{santos2022avoiding} or data transformations~\cite{devries2017improved,cubuk2020randaugment}. While simple, they only partially limit membership leakage, as they do not fundamentally eliminate the distinguishable patterns, and may hinder model convergence and degrade generative performance.

Differential privacy (DP) injects noise into the learning process to protect training data~\cite{dwork2008differential}, e.g., differentially private stochastic gradient descent (DP-SGD)~\cite{abadi2016deep}. It can defend against MIA in various deep learning models, however injecting noise leads to substantial degradation in generation quality and significantly increased computational cost, especially in diffusion models~\cite{truong2025attacks}.

Knowledge distillation is an alternative approach to achieve a better privacy–utility balance. It aims to distill unsensitive knowledge, e.g., identifiable samples, from the unprotected model trained on original dataset, and use it to train a privacy-preserved model~\cite{fernandez2023privacy}. It improves the trade-off but is computationally expensive and less scalable.

Adversarial training-based methods work through incorporating adversarial privacy objective during training~\cite{yan2025universal}. But current method is designed for fine-tuning~\cite{sun2021adversarial}, hence protecting only the privacy of the fine-tuning dataset. Additionally, it often introduces considerable training complexity.

Despite recent progress, existing defense methods for diffusion models still face limited effectiveness, privacy–utility trade-offs, and high computational overhead, underscoring the need for a more practical and efficient diffusion models-tailored defense mechanism with better trade-off between privacy, utility, and efficiency. 

\section{Preliminary and Problem Formulation}
\subsection{Diffusion Model}
Diffusion model learns data distribution and generates new samples by modeling a gradual transformation between origin data and Gaussian noise. It typically consists of a forward diffusion process and a reverse denoising process. In the forward process, a data sample $\mathbf{x}_0 \sim q(\mathbf{x})$ is progressively corrupted by Gaussian noise over $T$ timesteps:
\begin{equation}
\label{diffusion per step}
    q(\mathbf{x}_t | \mathbf{x}_{t-1}) = \mathcal{N}(\mathbf{x}_t; \sqrt{1 - \beta_t} \mathbf{x}_{t-1}, \beta_t \mathbf{I}),
\end{equation}
where $\{\beta_t\}_{t=1}^T$ is a predefined noise schedule. The whole forward process can be expressed as the cumulative effect of adding noise at each timestep:
\begin{equation}
    \label{diffusion total}
    q(\mathbf{x}_t | \mathbf{x}_0) = \mathcal{N}(\mathbf{x}_t;\sqrt{\overline{\alpha}_t}\mathbf{x}_0,(1-\overline{\alpha}_t)\mathbf{I}),
\end{equation}
where $\alpha_t=1-\beta_t$, $\overline{\alpha}_t=\prod_{i=1}^t\alpha_i$. Applying reparameterization trick on Equation ~\ref{diffusion total}: $\mathbf{x}_t = \sqrt{\overline{\alpha}_t}\mathbf{x}_0 + \sqrt{1-\overline{\alpha}_t}\boldsymbol{\epsilon}_t$, 
where $\epsilon_t \sim \mathcal{N}(0,\mathbf{I})$ is a sampled noise. Hence, any intermediate noised sample $\mathbf{x_t}$ can be sampled directly from $\mathbf{x_0}$.

The reverse process aims to recover the original data by training a deep neural network $\theta$ to learn the distribution $q(\mathbf{x}_{t-1} | \mathbf{x}_t)$:
\begin{equation}
    \label{denoising network}
    p_\theta(\mathbf{x}_{t-1} | \mathbf{x}_t)=\mathcal{N}(\mathbf{x}_{t-1};\mu_{\theta}(\mathbf{x}_t,t),\textstyle\sum_{\theta}(\mathbf{x}_t,t)),
\end{equation} 
where $\mu_{\theta}$ and $\textstyle\sum_{\theta}$ are the mean and variance parameterized by the neural network $\theta$.
The network is trained to accurately estimate the added noise $\epsilon_t$ at each timestep. Therefore, the training objective is to minimize the discrepancy between the predicted noise and the ground-truth noise, which leads to the following loss function:
\begin{equation}
    \label{loss function}
    \mathcal{L}_{\theta} = \mathbb{E}_{t, \mathbf{x}_0, \boldsymbol{\epsilon}} \left[ \| \boldsymbol{\epsilon} - \boldsymbol{\epsilon}_\theta(\mathbf{x}_t, t) \|^2 \right],
\end{equation}

\subsection{Membership Inference Attack}
Membership inference attacks (MIAs) aim to determine whether a given sample was included in the training dataset of the target model. Formally, let $\mathcal{D}$ denote a dataset, which can be partitioned into a training set $\mathcal{D}_{\text{train}}$ and a hold-out set $\mathcal{D}_{\text{hold}}$, where $\mathcal{D}_{\text{train}} \cap \mathcal{D}_{\text{hold}} = \emptyset$. A target model $f_\theta$ with parameters $\theta$ is trained on $\mathcal{D}_{\text{train}}$. The attacker can query the model with different access and obtain the output $f_{\theta}^{target}(x)$ from the model of a given query sample $x$ and determine the membership. MIA can be formulated as a binary classification task: $m(\mathbf{x}) = 1 \text{ if } \mathbf{x} \in \mathcal{D}_{\text{train}},\text{ 0 otherwise}$. 
The attacker's goal is to design an attack model $\mathcal{A}$ to infer the membership status:
\begin{equation}
    \mathcal{A}(\mathbf{x}_i, f_{\theta}) = \mathbb{I}\left(s(\mathbf{x}_i; f_{\theta}) \ge \tau \right),
\end{equation}
where $s(\mathbf{x}_i; f_{\theta})$ measures the likelihood of $\mathbf{x}_i$ being a training sample of the target model, and $\tau$ denotes a predefined threshold. $\mathcal{A}(\mathbf{x}_i,f_{\theta})=1$ refers that $\mathbf{x}_i \in \mathcal{D}_{train}$. In practice, $s(\mathbf{x}_i; f_{\theta})$ can be instantiated using various attack features derived from the model with different prior knowledge, such as generated likelihood, denoising estimated error, loss, gradient, etc.

\subsection{Problem Formulation}
In this work, we aim to design a defense framework for diffusion models to mitigate the risks posed by MIAs, where the adversary can exploit attack features based on model's overfitting. Formally, we consider a target diffusion model trained on $\mathcal{D}_{\mathrm{train}}$ and vulnerable to an attacker $\mathcal{A}$ that infers the membership of a query sample $\mathbf{x}$ according to the attack formulation introduced above. As discussed in the previous section, the success of such attacks stems from the model's over-memorization of training samples, which causes distinguishable behaviors between member and nonmember data during the denoising process.

Therefore, our objective is to learn a defended diffusion model that reduces the distinguishability exploited by MIAs while preserving the model's generative capability. Let $\theta$ denote the parameters of the denoising network, and let $\mathbf{M}$ denote a timestep-wise mask matrix that regulates the exposure pattern of training data over the standard diffusion training objective $\mathcal{L}_{\theta}$ in Eq.~(\ref{loss function}).

Based on the attack formulation above, we characterize the privacy objective as minimizing the membership inference risk of the defended model:
\begin{equation}
\label{eq:privacy_goal}
    \mathbf{Privacy \ goal:} \ 
    \min_{\theta,\mathbf{M}} \mathcal{R}_{\mathrm{MIA}}(\theta,\mathbf{M}),
\end{equation}
where
\begin{equation}
\label{eq:privacy_risk}
    \mathcal{R}_{\mathrm{MIA}}(\theta,\mathbf{M})
    =
    \mathbb{E}_{\mathbf{x}\sim \mathcal{D}_{\mathrm{train}}\cup \mathcal{D}_{\mathrm{hold}}}
    \Big[
        \mathbb{I}\big(
            \mathcal{A}(\mathbf{x}, f_{\theta,\mathbf{M}})=m(\mathbf{x})
        \big)
    \Big].
\end{equation}
Here, $m(\mathbf{x})$ is the ground-truth membership indicator defined in the previous section. A smaller $\mathcal{R}_{\mathrm{MIA}}(\theta,\mathbf{M})$ indicates that the defended model exposes less distinguishable membership information to the attacker.

Meanwhile, the defended model should preserve the utility of the original diffusion model. Since the generation ability of diffusion models is fundamentally established through the denoising learning objective $\mathcal{L}_{\theta}$, a defense should avoid excessively disturbing this learning process and achieve better generation performance. Let $\mathcal{U}(\theta)$ denote the utility of the defended model, measured by generation-quality metrics such as FID.
The utility goal is:
\begin{equation}
\label{eq:utility_goal}
    \mathbf{Utility \ goal:} \ \max_{\theta} \ \mathcal{U}(\theta),
\end{equation}
which is realized by minimizing the estimating error of the added noise in Eq.~\ref{loss function}.

Accordingly, the problem addressed in this paper can be formulated as learning a defended diffusion model parameterized by $\theta$ together with a timestep-wise mask matrix $\mathbf{M}$ such that the membership inference risk is minimized, while the utility remains bounded:
\begin{equation}
    \label{eq:overall_problem}
    \min_{\theta,\mathbf{M}} \ \mathcal{R}_{\mathrm{MIA}}(\theta,\mathbf{M})
    \quad
    \text{s.t.}
    \quad
    \mathcal{U}(\theta,\mathbf{M}) \ge \gamma,
\end{equation}
where $\gamma$ is the utility budgets, respectively. In practice, these quantities above are not optimized in a joint closed form. Instead, we optimize a defended variant of the standard denoising objective $\mathcal{L}_{\theta}$ under the control of $\mathbf{M}$, while the resulting privacy-utility trade-off is achieved through the design of $\mathbf{M}$ and its associated hyper-parameters.

\section{Methodology}
\subsection{Overview}

\begin{figure*}[!t]
    \centering
        \includegraphics[width=0.95\textwidth]{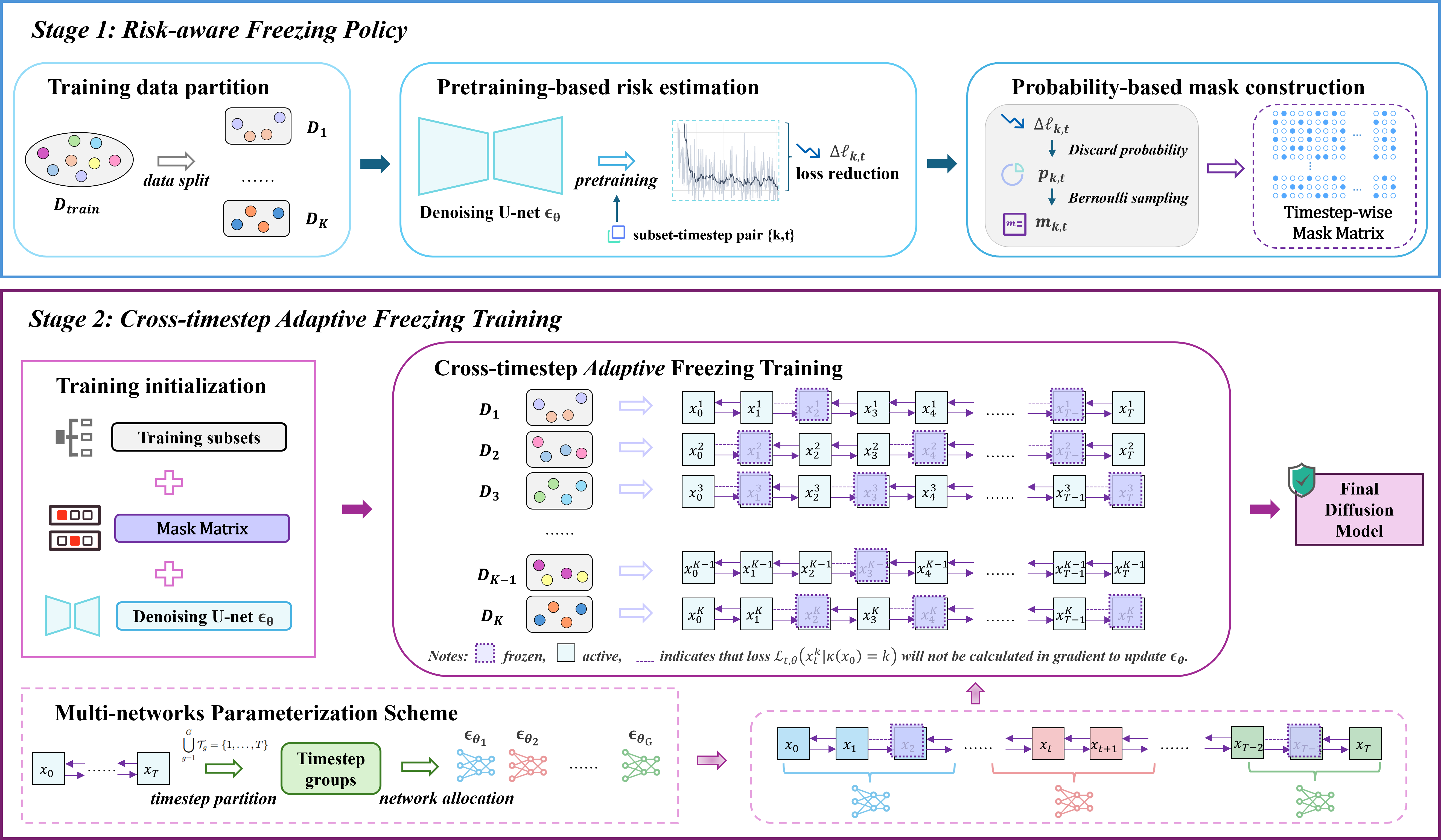}
        \caption{Adaptive Diffusion Freezing (ADF)}
        \label{fig:defense_framework}
\end{figure*}

MIAs against diffusion models exploit the membership-dependent discrepancies produced during multi-timestep denoising. Since each training sample repeatedly contributes across the denoising trajectory, over-memorization may vary across timesteps and data subsets, leading to distinguishable timestep-wise attack signals. This suggests that a defense should not uniformly suppress the entire training process; instead, it should adaptively reduce the influence of memorization-prone training signals while preserving the denoising dynamics necessary for high-quality generation.

Motivated by this observation, we propose \textbf{Adaptive Diffusion Freezing (ADF)} , a training-time defense framework that explicitly controls memorization at subset-timestep level, which we identify as a key source of membership inference risks. 
The main idea of ADF is to regulate over-memorization during training by adaptively controlling the appropriate contribution of different data subsets, thereby preventing memorization-prone samples from exerting disproportionate influence on the model at specific timesteps. Concretely, we introduce a timestep-wise mask matrix that determines whether a subset participates in optimization at a given timestep. Masked entries effectively block the loss contribution and gradient propagation, thereby freezing the influence of subset-timestep pairs with higher risk during training. 
ADF is built upon two key design mechanisms, as illustrated in Figure~\ref{fig:defense_framework}. First, we employ a cross-timestep adaptive freezing training strategy that partitions the training data into multiple subsets and controls their participation across timesteps, breaking the uniform exposure pattern inherent in diffusion training and reducing the contribution of specific samples with high memorization tendency.
Second, we design a risk-aware freezing policy that estimates memorization tendency based on training dynamics and adaptively suppresses high-risk training signals, enabling fine-grained control over the privacy–utility-efficiency trade-off.
By adaptively suppressing memorization-prone training subsets, ADF effectively mitigates overfitting-induced membership leakage risk while preserving the generative capability of diffusion models, achieving a favorable balance between privacy protection, utility, and efficiency.

\subsection{Cross-timestep Adaptive Freezing Training}
To mitigate model's over-memorization on training data, we need to regulate how training samples participate in diffusion process across timesteps. In standard diffusion training, each sample contributes repeatedly throughout the entire denoising trajectory. Such a uniform exposure pattern means that every part of the training data continuously affect optimization at all timesteps, which makes it difficult to explicitly limit memorization. Therefore, instead of allowing all training data to participate uniformly throughout the diffusion process, we seek a training mechanism that adaptively restricts the contribution of different parts of the training dataset at different timesteps.

Based on this intuition, our framework proposes cross-timestep adaptive freezing training, which consists of two parts: (1) Data splitting and mask matrix construction, and (2) design of freezing training. The details are demonstrated as follows.

\textbf{Data splitting and mask matrix construction.} We first partition the original training dataset $\mathcal{D}_{train}$ equally into $K$ non-overlapping parts, which are used as training subsets in ADF:
\begin{equation}
\label{dataset partition}
    \mathcal{D}_{train} = \bigcup_{k=1}^{K} \mathcal{D}_k, \quad \mathcal{D}_i \cap \mathcal{D}_j = \emptyset.
\end{equation}

Instead of being used to train separate sub-models, these training subsets serve as controllable units of training exposure across timesteps. Therefore, the training process can be regulated at the subset-level to gain fine-grained control over the training information. To associate each sample with its subset, we define a subset index function $\kappa(\mathbf{x}_0)\in\{1,\dots,K\}$, which returns the subset identity of a training sample $\mathbf{x}_0$.

Similarly to regular diffusion models, we define the forward and reverse processes as a Markov chain with $T$ timesteps. The forward process gradually adds Gaussian noise to the original image, while the reverse process learns a denoising network to predict the added noise and reconstruct the clean image. To apply adaptive freezing in diffusion setting, we introduce a binary mask matrix $\mathbf{M} \in \{0,1\}^{K \times T}$, where $T$ denotes the number of diffusion timesteps. Each element $\mathbf{m}_{k,t}$ specifies whether subset $\mathcal{D}_k$ participates in the optimization at timestep $t$ during the training process. Concretely, $\mathbf{m}_{k,t}=1$ means that samples from subset $\mathcal{D}_k$ are active at timestep $t$, whereas $\mathbf{m}_{k,t}=0$ means that their contribution will be frozen at that timestep. This mask matrix defines a timestep-wise exposure pattern over the entire training set. Rather than exposing every subset to every timestep, ADF allows different subsets to participate only at unfrozen timesteps, thereby providing a structured mechanism to regulate over-memorization throughout the diffusion trajectory.

\textbf{Design of freezing training.} Based on the mask matrix $\mathbf{M}$, we apply adaptive freezing to the reverse diffusion process. 
Following standard DDPM training~\cite{ho2020denoising}, for a clean sample $\mathbf{x}_0$, a timestep $t \sim \mathrm{Uniform}(\{1,\dots,T\})$, and Gaussian noise $\boldsymbol{\epsilon} \sim \mathcal{N}(\mathbf{0}, \mathbf{I})$, the standard denoising loss for the sample $\mathbf{x}_0 \in \mathcal{D}_k$ at timestep $t$ is:
\begin{equation}
\label{traditional loss}
    \mathcal{L}_{t,\theta}(\mathbf{x}_0)
        =
        \mathbb{E}_{\boldsymbol{\epsilon}}
        \left[
        \left\|
        \boldsymbol{\epsilon} - \boldsymbol{\epsilon}_{\theta}(\mathbf{x}_t, t)
        \right\|^2
        \right].
\end{equation}

To incorporate adaptive freezing, we use the mask entry corresponding to the sample subset and current timestep as a gating indicator:
\begin{equation}
\label{sample-mask}
    g(\mathbf{x}_0,t) = m_{\kappa(\mathbf{x}_0),t}.
\end{equation}
This indicator determines whether $\mathbf{x}_0$ is used in denoising training at timestep $t$. If $g(\mathbf{x}_0,t)=1$, the sample contributes normally to the denoising objective; if $g(\mathbf{x}_0,t)=0$, its loss is removed from the current mini-batch and no gradient is propagated from that sample at timestep $t$.

Formally, for a mini-batch $\mathcal{B}=\{\mathbf{x}_0^{(i)}\}_{i=1}^{|\mathcal{B}|}$, we define the active subset of samples at timestep $t$ as
\begin{equation}
\label{active-batch}
    \mathcal{B}^{\mathrm{act}}_{t}
    =
    \left\{
        \mathbf{x}_0^{(i)} \in \mathcal{B}
        \;\middle|\;
        m_{\kappa(\mathbf{x}_0^{(i)}),t}=1
    \right\}.
\end{equation}
Only samples in $\mathcal{B}^{\mathrm{act}}_{t}$ are used to compute the denoising objective. Accordingly, the freezing-aware mini-batch loss can be written as
\begin{equation}
\label{masked-batch-loss}
    \mathcal{L}_{\mathrm{ADF}}(\mathcal{B}, t)
    =
    \frac{1}{|\mathcal{B}^{\mathrm{act}}_{t}|}
    \sum_{\mathbf{x}_0^{(i)} \in \mathcal{B}^{\mathrm{act}}_{t}}
    \mathcal{L}_{t,\theta}(\mathbf{x}_0^{(i)}),
\end{equation}
where $|\mathcal{B}^{\mathrm{act}}_{t}|$ denotes the number of active samples in the current mini-batch at timestep $t$. Equivalently, this objective can be expressed in expectation form as
\begin{equation}
\label{masked loss}
    \mathcal{L}_{\mathrm{ADF}}
    =    \mathbb{E}_{t,\mathbf{x}_0,\boldsymbol{\epsilon}}
    \left[
        m_{\kappa(\mathbf{x}_0),t}
        \cdot
        \mathcal{L}_{t,\theta}(\mathbf{x}_0)
    \right].
\end{equation}

In this sense, ADF performs freezing training not by modifying the network architecture, but by adaptively removing the optimization signal of certain subset-timestep pairs. As shown in Figure~\ref{fig:defense_framework}, under this framework, different subsets contribute adaptively across timesteps, breaking the uniform training pattern and preventing consistent overfitting to specific data groups. As a result, the model's memorization behavior is reduced, which directly weakens the signals exploited by MIAs.

In addition, ADF suppresses over-memorization without indiscriminately weakening the overall learning capacity of the diffusion model. Unlike global regularization or noise injection, which uniformly perturb the entire training process and degrade denoising quality, ADF only freezes a small portion of subset-timestep pairs by blocking the loss contribution of selected pairs. This framework preserves the majority of informative training signals, allowing the model to continue learning the global data distribution and denoising dynamics from the remaining active subsets. Moreover, the mask matrix operates at a fine granularity: a subset frozen at one timestep can still contribute at other timesteps, while a timestep with some frozen subsets still receives supervision from the remaining active subsets. Therefore, ADF does not impair the model's ability to learn the global generation task, but only prevents certain samples from exerting excessive influence at specific timesteps. This adaptive rather than global suppression is the key to reducing memorization while maintaining generative quality.

\textbf{Multi-network Parameterization Scheme}
To further enhance the flexibility of ADF, we extend the framework to a multi-network parameterization scheme, where different timestep groups are modeled by separate denoising networks. The motivation is that diffusion timesteps correspond to different noise levels and denoising subtasks, and their memorization behavior as well as privacy sensitivity may vary substantially across the diffusion trajectory. When all timesteps share a single denoising network, the parameters for different denoising regimes are tightly coupled, which may limit the granularity of privacy control. By contrast, assigning different timestep groups to separate networks can decouple the learning dynamics of different denoising regimes and localize the effect of adaptive freezing within each timestep group, thereby providing more flexible control over the privacy-utility-efficiency trade-off.

Formally, we partition the $T$ diffusion timesteps into $G$ disjoint timestep groups:
\begin{equation}
\label{eq:timestep_group_partition}
    \bigcup_{g=1}^{G} \mathcal{T}_g = \{1,\ldots,T\},
    \qquad
    \mathcal{T}_g \cap \mathcal{T}_{g'} = \emptyset \ \text{for } g \neq g'.
\end{equation}
For each timestep group $\mathcal{T}_g$, we associate a dedicated denoising network $\epsilon_{\theta_g}$. Let $g(t)$ denote the group index such that $t \in \mathcal{T}_{g(t)}$. Then, for a training sample $\mathbf{x}_0 \in \mathcal{D}_k$, the masked denoising objective is extended as
\begin{equation}
\label{eq:multi_network_adf}
    \mathcal{L}_{\mathrm{ADF\text{-}MN}}
    =
    \mathbb{E}_{t,\mathbf{x}_0,\boldsymbol{\epsilon}}
    \left[
        m_{\kappa(\mathbf{x}_0),t}
        \cdot
        \left\|
            \boldsymbol{\epsilon}
            -
            \boldsymbol{\epsilon}_{\theta_{g(t)}}(\mathbf{x}_t,t)
        \right\|^2
    \right].
\end{equation}

This extension preserves the core idea of ADF: the mask matrix still determines whether a subset participates in optimization at each timestep, while the denoising network is selected according to the timestep group. Therefore, adaptive freezing continues to operate at the subset-timestep level, but the denoising process is parameterized more flexibly through timestep group-specific networks. In this way, the proposed extension can further reduce interference among different denoising regimes, improve modeling flexibility, and provide finer-grained privacy control while maintaining the denoising capability required for high-quality generation.

\subsection{Risk-aware Freezing Policy}
A key challenge in ADF is how to construct the mask matrix $\mathbf{M}$ in a principled manner. Although randomly initialized mask can realize freezing training, but a well-designed mask matrix can lead to better privacy-utility trade-off. Therefore, beyond adaptive freezing training, ADF requires a policy to identify where freezing should be applied to reduce unnecessarily harming generation quality.

A straightforward solution is to optimize $\mathbf{M}$ jointly with the diffusion model during training as part of the model parameters. Prior work on learnable masking has shown that mask variables can be incorporated into the training objective and progressively sparsified with $L_0$ regularization. In these settings, masks are typically attached to model activations or weights and remain inside the differentiable computation graph, so their updates can be guided by end-to-end gradients and sparsity regularization. For instance, Liu et al. \cite{liu2025masks} learn masks for the activation of experts in large language models to reduce computational load by initializing all units as active and progressively sparsifying them during training under an $L_0$ constraint, which is effective because the masked variables are directly coupled with the forward activations of the network.

However, applying such a design to the optimization of the mask matrix in diffusion setting is inappropriate. First, $\mathbf{M}$ does not gate hidden activations or model weight, instead, it controls whether a training subset participates in optimization at a given timestep. Once $m_{k,t}=0$, the corresponding samples are excluded from the loss computation at timestep $t$, and therefore do not contribute gradients to update either the diffusion model or mask entry. This creates a fundamentally different optimization problem from activation masking. As a result, jointly optimizing $\mathbf{M}$ with the model tends to be unstable: the learned mask may collapse toward overly sparse solutions driven by regularization and exhibit limited discriminative structure across timesteps and subsets. Moreover, because standard diffusion training is highly non-stationary across timesteps, the quality of training-time mask updates can be dominated by short-term optimization noise rather than the underlying memorization tendency. More importantly, online mask learning is conceptually misaligned with the privacy objective of ADF. In our framework, the mask matrix is intended to specify which subset-timestep pairs should be excluded before the main adaptive freezing stage begins, so that high-risk exposure can be prevented before it occurs. However, if the mask is optimized jointly with the diffusion model, some subsets may still participate in training at early stages before the corresponding mask entries are updated. Once such data have already contributed to parameter updates, the induced exposure cannot be undone by later switching the mask from 1 to 0. Therefore, for ADF, the mask should serve as a predefined constraint for the main adaptive freezing stage, rather than learned online after all subsets have been exposed to the model.

To address these issues, we decouple mask construction from the main adaptive freezing stage and introduce a pretraining process to construct a risk-aware freezing policy, as shown in Figure~\ref{fig:defense_framework}. The key idea is to first observe how the model fits different subsets at different timesteps under standard diffusion training, and then use these early fitting dynamics as a proxy for memorization tendency of subset-timestep pairs. Based on the estimated memorization tendency, the policy assigns higher freezing priority to pairs with higher risk.

\textbf{Pretraining-based risk estimation.}
To instantiate the risk-aware freezing policy, we need a proxy that reflects how strongly the model tends to remember different subset-timestep pairs. We adopt loss reduction as this risk signal, since memorization is closely related to how rapidly and effectively the model fits training data. Prior studies on membership inference have shown that privacy leakage fundamentally benefits from overfitting, and that samples that are fitted more easily or more confidently are more likely to expose distinguishable membership signals~\cite{shokri2017membership,yeom2018privacy}. In diffusion models, such fitting dynamics are naturally reflected by the denoising loss~\cite{duan2023diffusion}. Motivated by these observations, we use the reduction of denoising loss during early training as a proxy for memorization tendency. If the loss of a subset decreases substantially at a given timestep, it indicates that the model can absorb the corresponding training signal more aggressively there, which in turn suggests a stronger tendency toward over-memorization.
Therefore, rather than treating all subset-timestep pairs equally, we use loss reduction to estimate their relative level of memorization.

Specifically, we first perform $n$ epochs of standard diffusion training using all subsets (i.e., $\mathbf{M}$ initialized as an all-one matrix). During this phase, we record the denoising loss for each subset $k$ at timestep $t$ in epoch $i$: $\ell_{k,t}^{(i)}$. We then quantify the learning dynamics via loss reduction:
\begin{equation}
\label{pretraining loss reduction}
    \Delta \ell_{k,t} = \ell_{k,t}^{(1)} - \ell_{k,t}^{(n)},
\end{equation}
which reflects how effectively the model fits subset $\mathcal{D}_k$ at timestep $t$. A larger $\Delta \ell_{k,t}$ indicates stronger fitting and a higher risk of memorization.

\textbf{Probability-based mask construction.} 
The loss reduction $\Delta \ell_{k,t}$ serves as a continuous risk score, but the mask matrix itself is binary. Therefore, we first convert the estimated memorization tendency into a freezing probability: 
\begin{equation}
\label{freezing probability}
    p_{k,t} = \text{clip}_{\alpha}^{\beta} \left( \sigma(\Delta \ell_{k,t}) \right),
\end{equation}
where $\sigma(\cdot)$ is the sigmoid function, and $\alpha, \beta$ are lower and upper bounds to stabilize the probability range. This transformation has two purposes: First, it maps the raw loss reduction, whose scale can vary between subset-timestep pairs, to a normalized and comparable quantity, so that larger $\Delta \ell_{k,t}$ consistently corresponds to a higher freezing tendency. Second, it avoids directly imposing a hard threshold on $\Delta \ell_{k,t}$, which would make the mask construction overly sensitive to local fluctuations in the estimated training dynamics. 

Based on the freezing probability, the mask matrix is then generated from a Bernoulli sampling with activation probability of $1 - p_{k,t}$. In this way, subset-timestep pairs with larger loss reduction are more likely to be frozen, while pairs with lower reduction are more likely to remain active.

\textbf{Sparsity constraint.}  
While the probability-based masking strategy enables ADF to prioritize subset-timestep pairs with higher memorization tendency, applying freezing without any global control may result in an overly aggressive mask matrix. In particular, if too many entries are set to zero, the defended model would lose a substantial portion of denoising supervision at certain timesteps, which may impair the continuity of diffusion learning and significantly degrade generative quality. Therefore, beyond identifying where freezing should be applied, ADF must also explicitly control how much freezing is allowed. 

To this end, we impose a hard constraint on each timestep: $\sum_{k=1}^{K} (1 - m_{k,t}) = s \cdot K$, 
where $s \in [0,1]$ is a sparsity hyperparameter that determines the fraction of frozen subsets at timestep $t$. Equivalently, each timestep retains $(1-s)K$ active subsets to participate in denoising optimization.

The sparsity constraint is crucial for maintaining generative utility. Diffusion models require sufficient supervision across the denoising trajectory in order to learn stable transitions between neighboring timesteps. If freezing is left unconstrained, high-risk subsets may be removed excessively at some timesteps, causing the model to underfit the corresponding denoising subtask and weakening the overall generation process. By enforcing a fixed sparsity budget, ADF ensures that only a limited portion of the training exposure is suppressed, while the majority of informative training signals remain available. As a result, the model can still learn the global data distribution and timestep-wise denoising dynamics from the retained active subsets. 

Moreover, the sparsity constraint serves as an explicit control knob for the privacy-utility trade-off. A larger $s$ increases the amount of freezing and thus tends to provide stronger suppression of over-memorization, but it also reduces the amount of effective supervision and may harm generation quality. In contrast, a smaller $s$ preserves more denoising signals and better maintains utility, but may leave more memorization-related signals in training. Therefore, $s$ directly controls the balance between privacy protection and generative performance, which enables ADF to preserve generative utility while still achieving effective MIA defense. 


\section{Experiments}
\subsection{Experiment Setup}
\subsubsection{Datasets} 
We evaluate our method on four widely used image datasets covering natural images, facial images, and remote sensing imagery.

\textit{CIFAR-10}~\cite{krizhevsky2009learning}. 
CIFAR-10 is a benchmark dataset in computer vision. It contains 60,000 color images of size $32 \times 32$ from 10 classes, with 50,000 training and 10,000 testing samples.
For our experiments, we follow the original split and randomly select 50\% of the training set to train the target model and use the remaining 50\% as the hold-out set for membership inference evaluation.


\textit{STL10\_U}~\cite{coates2011analysis}. 
STL10\_U is the unlabeled split of the STL-10 dataset, which contains 100,000 natural images with a resolution of $96 \times 96$. Compared with CIFAR-10, STL10\_U contains higher-resolution images and more diverse visual content. The data split is same as CIFAR-10.

\textit{CelebA}~\cite{liu2015deep}. 
CelebA is a large-scale face dataset containing over 200,000 images annotated with 40 binary attributes and five facial landmarks. The dataset includes substantial variations in identity, pose, and illumination, and is widely used for generative modeling tasks. We follow the same data split protocol as CIFAR-10.

\textit{NWPU-RESISC45}~\cite{cheng2017remote}. 
NWPU-RESISC45 is a remote sensing scene classification dataset consisting of 31,500 images from 45 scene classes, with 700 images per class. The images are collected from low-altitude aerial and satellite platforms and cover diverse ground scenes such as airport, basketball court, bridge, lake, parking lot, etc. We randomly partition it into two disjoint equal subsets, i.e., a training set and a hold-out set, while approximately preserving class balance.

\subsubsection{Baselines}
We redapply five baselines, including a diffusion model without any defense and four representative defense strategies. Specifically, the five baselines are as follows: (1) W/O: we use standard diffusion model as the undefended baseline. The main experiments instantiate it with DDPM, while the applicability of ADF to other diffusion architectures is examined subsequently. (2) Cutout: we follow the standard Cutout augmentation strategy~\cite{devries2017improved} and randomly masks out contiguous square regions of the input image with a single-hole Cutout operation with a mask size of $8 \times 8$ before normalization during training. (3) L2: following~\cite{van2017l2}, we apply L2 regularization to diffusion training by incorporating a weight decay term into the Adam optimizer. (4) DP-SGD: we follow ~\cite{abadi2016deep} and perturb gradient updates with calibrated noise during training. (5) Knowledge distillation (KD): following the privacy-distillation paradigm in~\cite{fernandez2023privacy}, we use a teacher diffusion model to generate and filter synthetic samples based on the nearest-neighbor distances from real training data, which indicating potential privacy risks. The distilled student diffusion model, trained solely on filtered synthetic samples, is used for final evaluation.

\subsubsection{MIA Methods}
To evaluate our proposed framework under multiple MIA settings with different adversary capabilities, we consider MIAs with two representative threat models. Specifically, we adopt SecMI~\cite{duan2023diffusion} and GSA~\cite{pang2023white}, which correspond to gray-box and white-box attacks, respectively.
SecMI assumes that the adversary can access denoising-related intermediate information, i.e., loss and noisy image, from the diffusion process and infer membership. Following the original design, we report results for both its statistical variant and neural-network-based variant, denoted as $SecMI_{stat}$ and $SecMI_{nn}$. In contrast, $GSA$ assumes stronger adversary capabilities, including access to model gradients and internal parameters, and thus provides a more stringent white-box evaluation of privacy leakage. Since these attacks exploit the characteristic discrepancies between member and nonmember samples from different perspectives and under strong assumptions of the adversary capability, defending against them provides a rigorous assessment of the defense effectiveness.

\subsubsection{Evaluation Metrics} We use the following metrics to measure the defense performance, utility and efficiency of the defense model. We employ four common metrics to evaluate the defense performance against MIA, i.e. Attack success rate (ASR), AUC, TPR@1\%FPR, and TPR@0.1\%FPR. These metrics measure the overall inference correctness, discriminate ability, and attack effectiveness at low false-positive rates of MIA, respectively. A lower metric value indicates better privacy protection. We adapt FID to measure the fidelity and quality of the synthetic data, which reflect the utility of a generative model. We also report training time and trainable parameters, which reflect the computational efficiency of the defense model.

\subsubsection{Implementation Details}
We adopted a DDPM-style diffusion backbone with a U-Net denoiser and trained all models with $T=1000$ diffusion steps. For fair comparison, we used the same backbone architecture, image resolution, and optimization pipeline across the undefended model and all defended variants, and only changed the defense-specific components. Unless otherwise stated, all models were trained with the same noise schedule, optimizer, and sampling protocol. For ADF, unless otherwise specified, we partitioned the training data into $K=10$ subsets, used $G=1$ timestep group, set sparsity to $s=0.3$, and performed risk estimation with $5$ pretraining epochs.
All experiments were conducted on two servers. The primary server ran an Ubuntu 22.04 with an Intel(R) Xeon(R) Gold 6459 CPU, 90 GB RAM, and a NVIDIA GeForce RTX 5090 GPU with 32 GB memory. We used Python 3.12.3 and PyTorch 2.7.0 with CUDA 12.8 on this server. Additional experiments were conducted on a second server running Ubuntu 18.04.6 LTS and was equipped with dual Intel(R) Xeon(R) Gold 5320 CPUs, 119 GB RAM, and two NVIDIA A100-SXM4-80GB GPUs.

\subsection{Experimental Results}
\begin{figure*}[!t]
    \centering
    \begin{subfigure}[b]{\textwidth}
        \centering
        \includegraphics[width=0.95\textwidth]{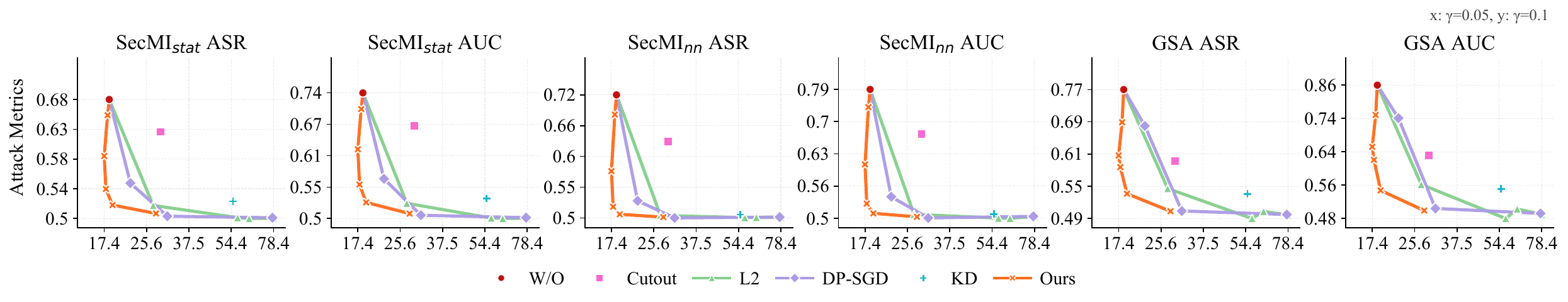}
        \caption{CIFAR-10}
        \label{fig:privacy_utility_tradeoff_cifar10}
    \end{subfigure}

    \begin{subfigure}[b]{\textwidth}
        \centering
        \includegraphics[width=0.95\textwidth]{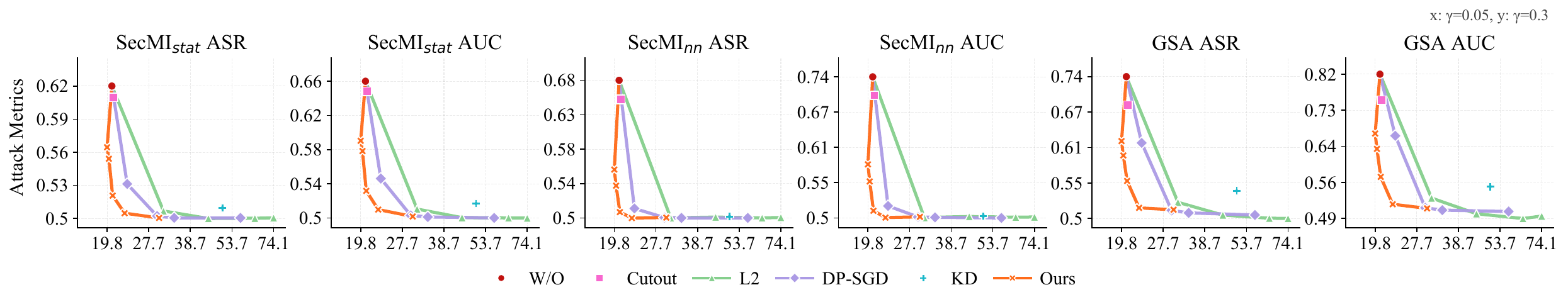}
        \caption{STL10\_U}
        \label{fig:privacy_utility_tradeoff_STL10U}
    \end{subfigure}

    \begin{subfigure}[b]{\textwidth}
        \centering
        \includegraphics[width=0.95\textwidth]{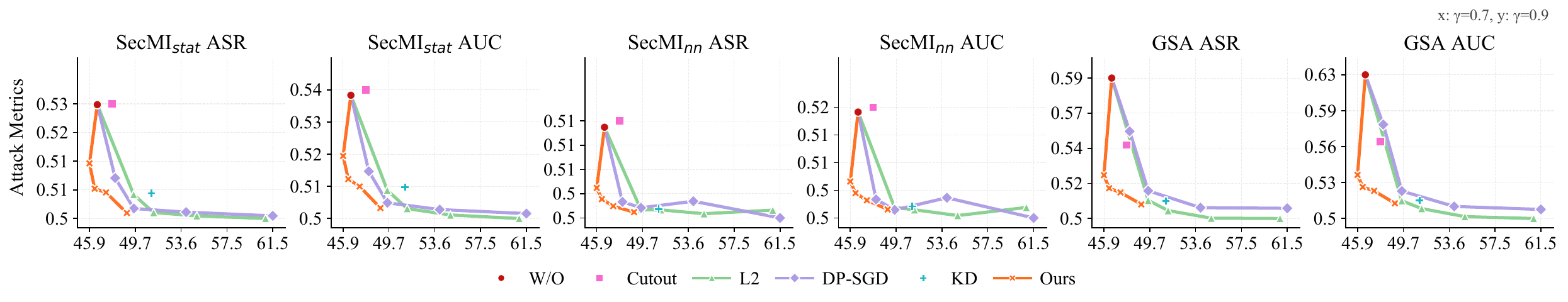}
        \caption{CELEBA}
        \label{fig:privacy_utility_tradeoff_celeba}
    \end{subfigure}

    \begin{subfigure}[b]{\textwidth}
        \centering
        \includegraphics[width=0.95\textwidth]{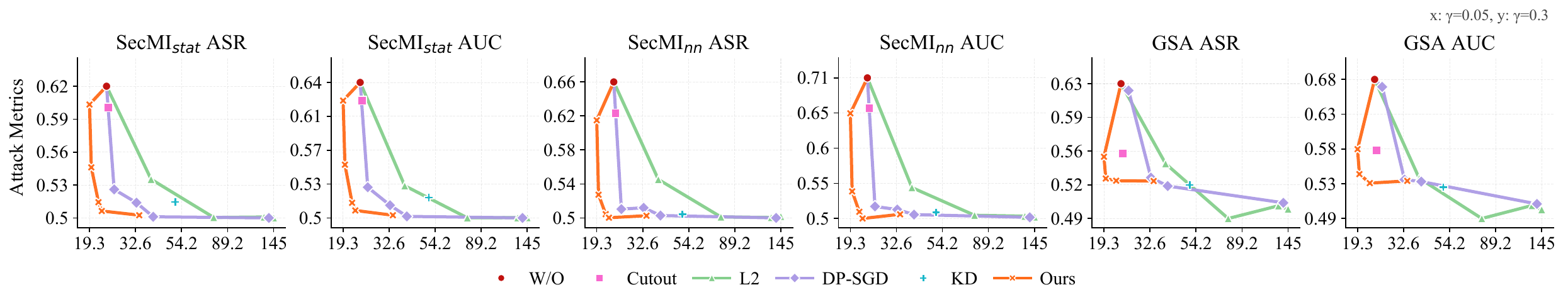}
        \caption{NWPU-RESISC45}
        \label{fig:privacy_utility_tradeoff_NWPU}
    \end{subfigure}
    \caption{Privacy-utility trade-off performance of ADF on DDPM compared with baselines. X-axis: FID of the target model; Y-axis: ASR or AUC of MIA attacks. Points closer to the lower-left corner indicate a better privacy-utility trade-off.}
    \label{fig:privacy_utility_tradeoff}
\end{figure*}

\subsubsection{Privacy-utility Trade-off Performance}
We compared the performance of privacy-utility trade-off of ADF with five baselines under three MIA methods and utility evaluation. We employ $SecMI_{stat}$, $SecMI_{nn}$, and $GSA$ as attack methods, and use ASR and AUC to measure membership leakage risk, which reflects the level of privacy preservation. 
In this setting, a lower value on the y-axis indicates stronger privacy protection, i.e., lower attack performance, while a lower value on the x-axis indicates better generative utility. Therefore, methods closer to the lower-left region achieve a better privacy-utility trade-off. By varying the defense-related hyper-parameters, including the weight decay coefficient for L2 regularization, the noise budget for DP-SGD, and the sparsity level for ADF, we obtain the corresponding trade-off curves. For methods without tunable hyper-parameters, including the undefended DDPM, Cutout, and knowledge distillation, we report them as individual points.

As shown in Figure~\ref{fig:privacy_utility_tradeoff}, ADF consistently achieves a superior privacy-utility trade-off across datasets and attack settings. The undefended DDPM usually achieves the best or near-best FID, but exhibits high MIA performance. It is located in the upper-left region of most plots. This confirms that diffusion models can expose distinguishable member-specific behaviors even when they generate high-quality samples. Cutout slightly reduces MIA performance in some cases, but its improvement over the undefended model is limited, suggesting that standard data augmentation cannot effectively suppress diffusion-specific, timestep-wise membership signals. In addition, knowledge distillation (KD) achieves stronger reductions in ASR and AUC, but at the cost of substantially degraded generation quality, as reflected by its position on the right side of the plots. Therefore, although KD can mitigate membership leakage, its utility loss limits its practicality in applications requiring high-fidelity.

Regarding the performance of privacy-utility trade-off, tunable defenses show a common trend: as the defense strength increases, the membership leakage risk decreases, as reflected by lower ASR or AUC, while the generative utility is gradually degraded, as reflected by higher FID values. As the regularization weight, noise level, or sparsity increases, the corresponding curves generally move downward, indicating lower attack performance, while shifting rightward to varying degrees due to FID degradation. More importantly, we observe that ADF consistently traces a more favorable trade-off frontier across datasets and MIA attacks. Its curve drops rapidly while remaining close to the left side of the plots, showing that ADF can substantially reduce ASR and AUC with only a small, or even negligible, increase in FID. Compared with L2 regularization and DP-SGD, ADF is consistently closer to the lower-left region. This indicates that, under the same privacy protection level, ADF preserves better generative utility; conversely, under comparable FID, ADF achieves lower membership leakage. Since ADF does not globally perturb or constrain the entire diffusion model, by increasing the sparsity level, it adaptively reduces the contribution of high-risk sample-timestep pairs. This allows the model to suppress timestep-wise membership leakage while retaining useful denoising knowledge from low-risk regions. As a result, ADF avoids the excessive utility degradation commonly observed in global defenses. 

In addition, we also discover a slightly increase on generative utility in ADF compared to DDPM. This suggests that moderately removing some training samples with high MIA risk may benefit the diffusion model. High-risk subset-timestep pairs are more likely to contain membership-specific memorization signals, which may also correspond to overfitted or less generalizable training behavior. Therefore, moderately suppressing these pairs can reduce overfitting and improve the learned denoising distribution. This observation further indicates that privacy preservation and generative utility are not always strictly conflicting in ADF: when the defense precisely targets memorization-prone components, it can reduce privacy leakage without sacrificing, and sometimes even improving, sample quality.


\subsubsection{Generalizability of ADF across Mainstream Diffusion Architectures}
We conduct experiments on different diffusion architectures, i.e., latent diffusion model (LDM) and multi-modal diffusion transformer (MMDiT). Specifically, we follow the standard latent diffusion paradigm and train a LDM locally on all experimental datasets. For MMDiT, we adopt the pretrained Stable Diffusion 3 (SD3) Medium from Hugging Face as the backbone and fine-tune over COCO2017-Val.

\begin{figure*}[!t]
    \centering
    \begin{subfigure}[b]{\textwidth}
        \centering
        \includegraphics[width=0.95\textwidth]{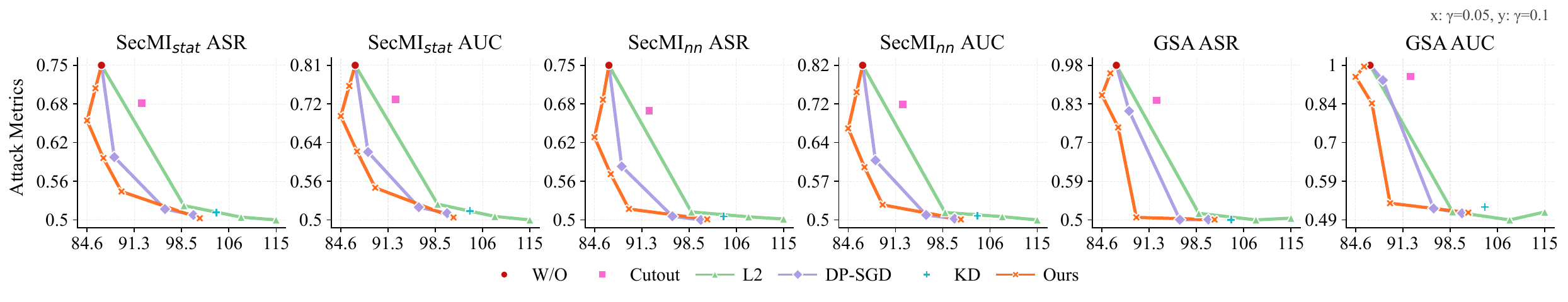}
        \caption{CIFAR-10}
        \label{fig:ldm_privacy_utility_tradeoff_cifar10}
    \end{subfigure}

    \begin{subfigure}[b]{\textwidth}
        \centering
        \includegraphics[width=0.95\textwidth]{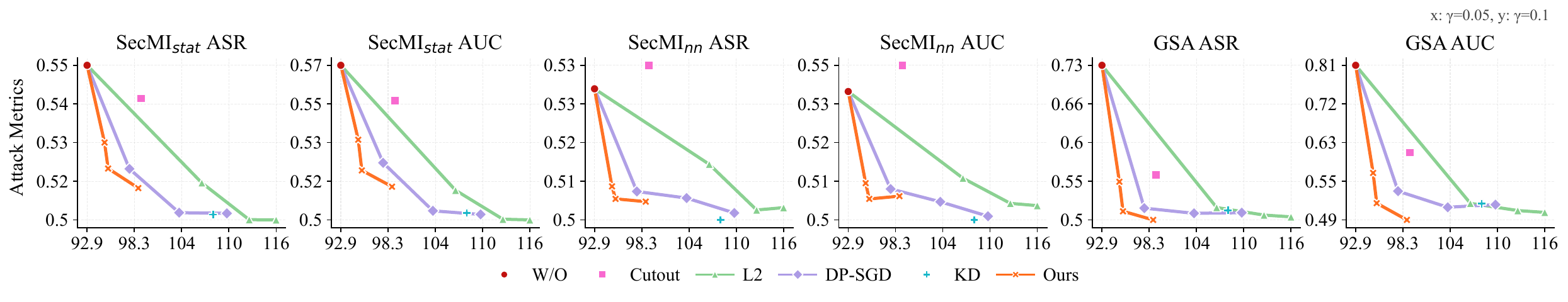}
        \caption{STL10\_U}
        \label{fig:ldm_privacy_utility_tradeoff_STL10U}
    \end{subfigure}

    \begin{subfigure}[b]{\textwidth}
        \centering
        \includegraphics[width=0.95\textwidth]{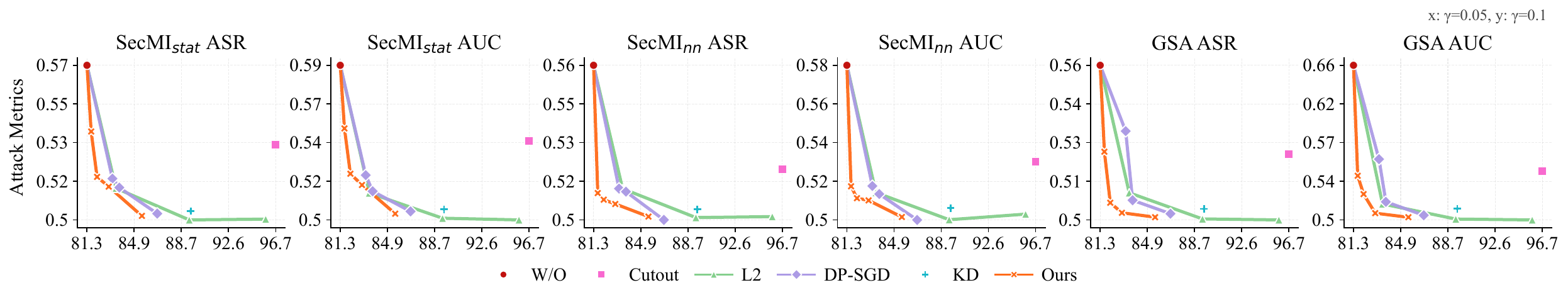}
        \caption{CELEBA}
        \label{fig:ldm_privacy_utility_tradeoff_celeba}
    \end{subfigure}

    \begin{subfigure}[b]{\textwidth}
        \centering
        \includegraphics[width=0.95\textwidth]{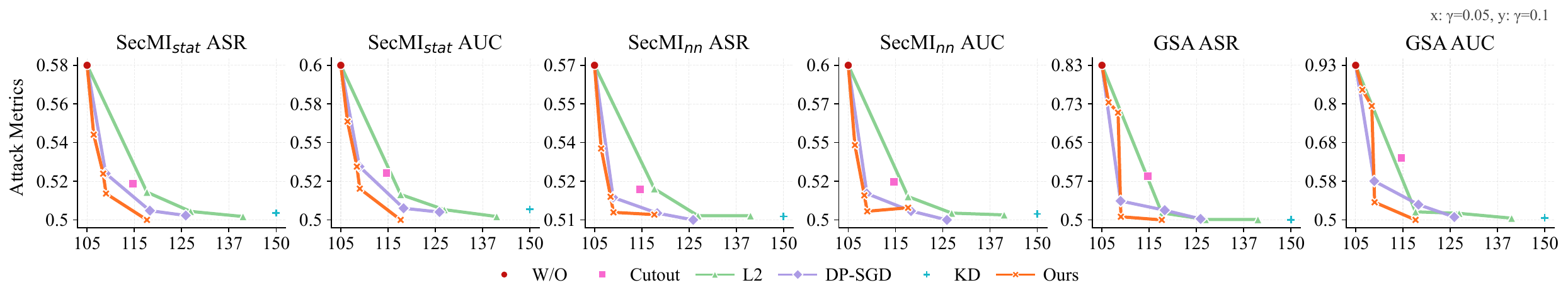}
        \caption{NWPU-RESISC45}
        \label{fig:ldm_privacy_utility_tradeoff_NWPU}
    \end{subfigure}
    \caption{Privacy-utility trade-off performance of ADF on LDM compared with baselines. X-axis: FID of the target model; Y-axis: ASR or AUC of MIA attacks. Points closer to the lower-left corner indicate a better privacy-utility trade-off.}
    \label{fig:ldm_privacy_utility_tradeoff}
\end{figure*}

Figure~\ref{fig:ldm_privacy_utility_tradeoff} presents the privacy-utility trade-off of ADF compared to various defense baselines, similar to the main experiment protocol. Across the evaluated datasets and MIA methods, ADF significantly lowers the attack metrics while maintaining competitive generation quality compared to the undefended LDM, and provides better trade-off performance between privacy and utility compared to other defense methods. These results demonstrate that the effectiveness of ADF is not restricted to pixel-space diffusion models and can be readily transferred to latent-space diffusion.

We further evaluate ADF on SD3-Medium to examine its applicability to MMDiT-style architectures. As shown in Table~\ref{tab:sd3_results}, ADF consistently reduces the attack performance of all three MIAs, while slightly improving the FID from 31.223 to 26.479. Therefore, ADF remains effective on the MMDiT-style architecture and provides improved membership privacy without sacrificing generation quality, supporting its generalizability across diffusion architectures.

\begin{table}[t]
\centering
\caption{Evaluation of ADF on Stable Diffusion 3.}
\label{tab:sd3_results}
\resizebox{\columnwidth}{!}{
\begin{tabular}{lccccccc}
\toprule
& & \multicolumn{2}{c}{SecMI$_{stat}$}
& \multicolumn{2}{c}{SecMI$_{nn}$}
& \multicolumn{2}{c}{$GSA$} \\
\cmidrule(lr){3-4}
\cmidrule(lr){5-6}
\cmidrule(lr){7-8}
\textbf{Model} & \textbf{FID}
& \textbf{ASR} & \textbf{AUC}
& \textbf{ASR} & \textbf{AUC}
& \textbf{ASR} & \textbf{AUC} \\
\midrule
SD3
& 31.223
& 0.520 & 0.519
& 0.519 & 0.516
& 0.627 & 0.682 \\

ADF
& \textbf{26.479}
& \textbf{0.517} & \textbf{0.516}
& \textbf{0.511} & \textbf{0.501}
& \textbf{0.601} & \textbf{0.653} \\
\bottomrule
\end{tabular}
}
\end{table}

\subsubsection{Sensitivity Analysis}
In this section, we further examine robustness of ADF and analyze how its key design choices affect privacy preservation and generative utility. 

\begin{figure}[h]
    \centering
    \begin{subfigure}[b]{\linewidth}
        \centering
        \includegraphics[width=0.96\linewidth]{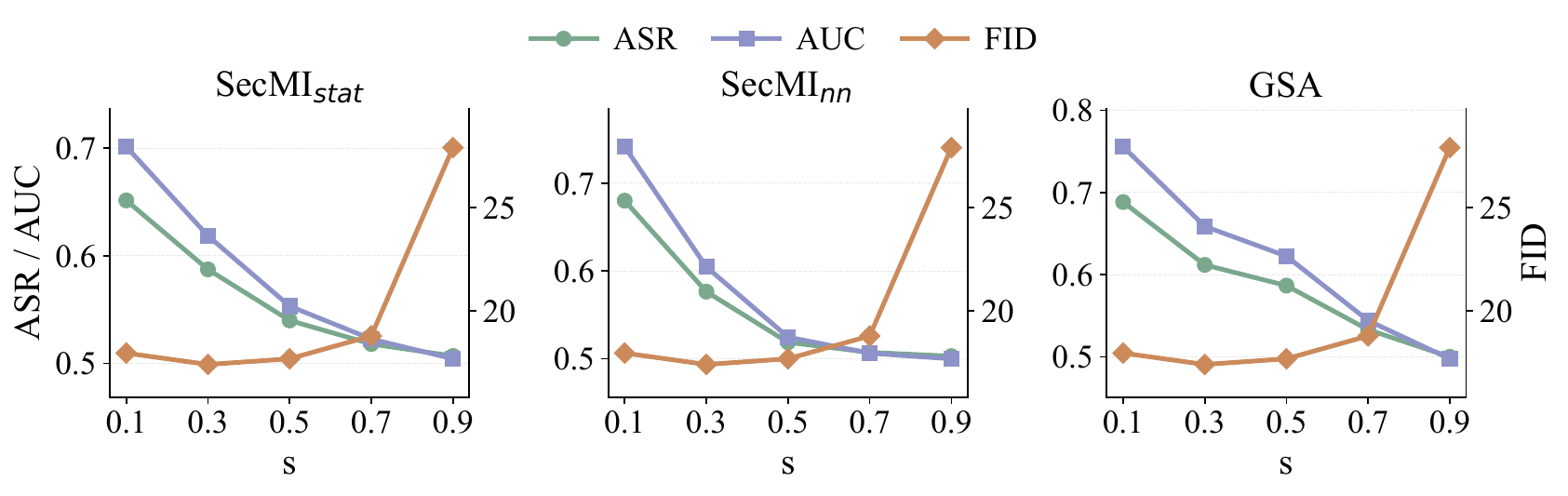}
        \caption{sparsity: s}
        \label{fig:ablation_s}
    \end{subfigure}
    \begin{subfigure}[b]{\linewidth}
        \centering
        \includegraphics[width=0.96\linewidth]{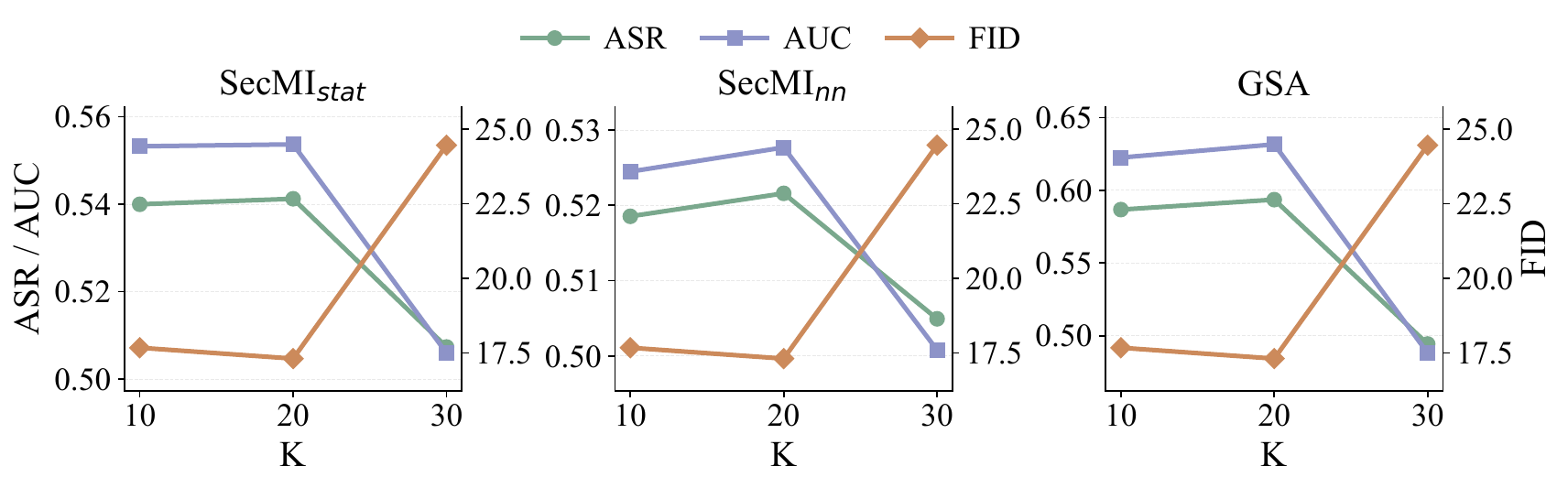}
        \caption{training subsets: K}
        \label{fig:ablation_k}
    \end{subfigure}
    \begin{subfigure}[b]{\linewidth}
        \centering
        \includegraphics[width=0.96\linewidth]{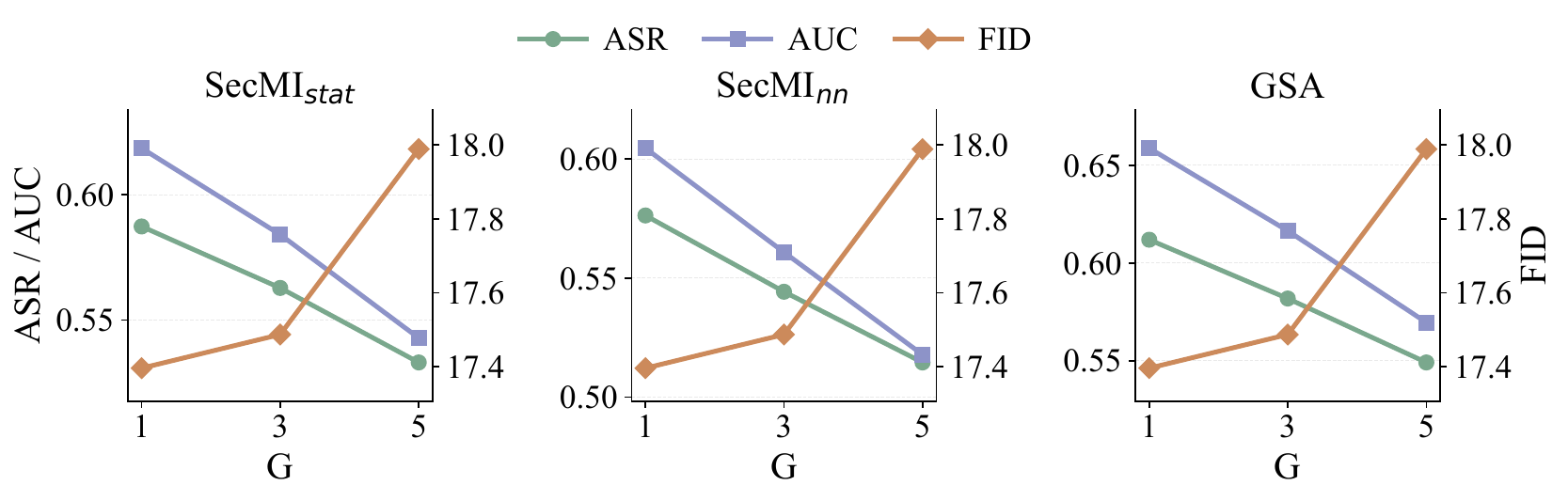}
        \caption{timestep group: G}
        \label{fig:ablation_g}
    \end{subfigure}
    \caption{Effect of hyper-parameters on privacy-utility trade-off performance.}
    \label{fig:ablation_k_s_g}
\end{figure}

\textbf{Effect of Mask Sparsity.}
Mask sparsity directly controls the strength of freezing. A larger sparsity means that more subset-timestep pairs are suppressed during training. We report its effect on privacy-utility trade-off in Figure~\ref{fig:ablation_s}. Increasing $s$ consistently reduces attack performance across $SecMI_{stat}$, $SecMI_{nn}$, and $GSA$. When $s$ is small, ADF only freezes a limited number of high-risk subsets at each timestep, so the model still over-memorize most of the training signals from the training samples and retains higher membership leakage. As $s$ increases, more risky subsets are removed, and the attack metrics gradually decrease toward random guessing. 
The improvement in privacy is accompanied by a gradual degradation in generation quality. FID remains relatively stable when $s$ increases from 0.1 to 0.5, indicating that ADF can suppress a considerable portion of membership leakage without heavily damaging utility. However, when $s$ becomes very large, especially at $s=0.9$, FID increases substantially. This shows that overly aggressive freezing removes not only privacy-sensitive memorization but also useful training signals required for high-quality denoising. Therefore, sparsity is the main knob for controlling the privacy-utility trade-off in ADF. A moderate sparsity level provides a favorable balance, while an extreme sparsity setting should be used only when privacy protection is prioritized over generation quality.


\textbf{Effect of the Number of Dataset Subsets.}
We study the effect of the number of dataset subsets $K$ under sparsity $s=0.5$, which determines the granularity of training sample partitioning in ADF.
Figure~\ref{fig:ablation_k} show that increasing $K$ from 10 to 20 does not significantly reduce ASR or AUC, but slightly improves FID, suggesting that moderate partitioning can preserve or even stabilize the denoising training process. However, when $K$ increases to 30, the attack performance drops sharply across all three MIAs, while FID increases noticeably. Therefore, finer partitioning helps ADF identify and freeze high-risk subset-timestep pairs more precisely, but overly fine partitioning may remove useful training signals and degrade generation quality. Therefore, the effect of $K$ reflects a non-linear privacy-utility trade-off rather than a simple monotonic improvement.


\textbf{Effect of the Number of Timestep Groups.}
We further evaluate the effect of the number of timestep groups $G$. This parameter controls how finely ADF models the timestep dimension. When $G=1$, all timesteps share the same freezing behavior, which degenerates ADF toward a coarse, model-level defense. As $G$ increases, ADF can assign different freezing decisions to different denoising stages and better capture the heterogeneity of membership leakage across timesteps.
Figure~\ref{fig:ablation_g} shows that increasing $G$ improves privacy protection under all three attacks. Compared with $G=1$, using multiple timestep groups reduces both ASR and AUC, confirming that timestep-wise modeling is important for suppressing diffusion-specific membership signals. This result is consistent with the observation that existing MIAs exploit timestep-dependent denoising errors, likelihoods, or gradients. 
The utility cost of increasing $G$ is relatively mild compared with the effect of increasing sparsity. FID increases only slightly from 17.396 to 17.989 as $G$ grows from 1 to 5. This indicates that finer timestep grouping mainly improves the precision of freezing rather than simply weakening the whole training process. However, applying multiple networks to each timestep group introduces higher computational costs, i.e., training time and storage space. Overall, the results validate a central design of ADF: explicitly modeling the timestep dimension is beneficial for privacy protection while only slightly causing generation degradation.


\begin{figure}[h]
    \centering
    \includegraphics[width=0.96\linewidth]{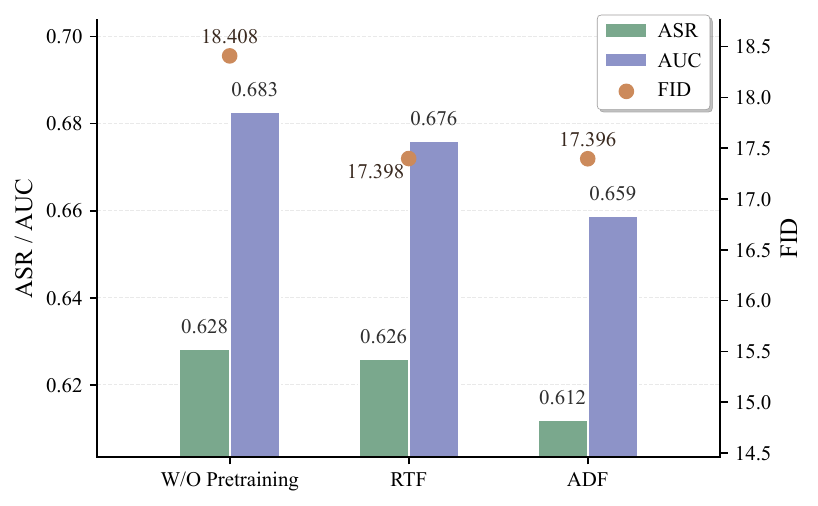}
    \caption{Ablation of risk-aware freezing policy on CIFAR-10 against $GSA$.}
    \label{fig:ablation_policy}
\end{figure}

\textbf{Effect of Risk-aware Freezing Policy.}
We examine the contribution of risk-aware freezing policy in ADF, comparing the full ADF with two variants: (1) a variant without pretraining-based risk estimation; and (2) an exposure-matched baseline, termed random timestep freezing (RTF). For each training sample, RTF randomly freezes an $s$ fraction of timesteps without using either the risk-aware policy or the subset structure. Therefore, each sample is trained on the same expected fraction of timesteps as the sparsity in ADF, ensuring a matched training-exposure budget, allowing us to distinguish the benefit of targeted freezing from that of simply reducing training exposure.
As shown in Figure~\ref{fig:ablation_policy}, removing the risk-aware policy leads to consistently higher attack performance and worse FID in the W/O pretraining variant. This demonstrates that the risk-aware policy is important to identify and suppress privacy-sensitive subset-timestep contributions while preserving useful denoising information.
Moreover, under the same expected training-exposure budget, ADF further reduces ASR/AUC while maintaining nearly identical generation quality to RTF. This comparison shows that the privacy improvement of ADF cannot be explained merely by reduced sample-level training exposure; rather, risk-aware allocation of the freezing budget toward more privacy-sensitive subset-timestep pairs provides additional protection without sacrificing generation quality.
We also observe that RTF achieves a better FID than the variant without pretraining. In the absence of risk-aware policy, the sample-wise random timestep freezing in RTF distributes training contributions more uniformly across samples, whereas random freezing at the subset-timestep level may unevenly suppress useful denoising updates and thus cause greater degradation in generation quality.
Overall, these results confirm that the effectiveness of ADF comes from the combination of two components: cross-timestep adaptive freezing and risk-aware freezing policy.


\textbf{Robustness across MIA Settings.}
We verify the defense robustness of ADF across various MIA methods and metrics. Based on the privacy-utility trade-off curves in Figure~\ref{fig:privacy_utility_tradeoff}, we select a representative operating point for each tunable defense whose FID is close to that of the undefended DDPM. This enables a fair comparison under comparable generative utility, rather than only comparing the strongest privacy settings. As illustrated in Table~\ref{tab:privacy_protection_performance} in appendix, ADF achieves a more balanced performance than the baselines: it substantially reduces membership leakage while maintaining FID comparable to or lower than the undefended model on most datasets. 
Across different attack methods and datasets, ADF shows robust protection against MIAs. Compared with the undefended model, ADF consistently lowers ASR and AUC under $SecMI_{stat}$, $SecMI_{nn}$, and $GSA$, indicating that member and nonmember samples become less distinguishable from multiple perspectives of attack features. In addition, ADF also lowers TPR under low-FPR regimes, i.e., TPR@1\%FPR and TPR@0.1\%FPR, which measure whether attackers can still identify members under strict false-positive constraints. These reductions provide a stricter evaluation of privacy protection. 
The consistent decrease of these TPR at low-FPR values shows that ADF not only reduces general MIA accuracy, but also suppresses highly exposed membership signals.

\begin{figure}[h]
    \centering
    \includegraphics[width=0.96\linewidth]{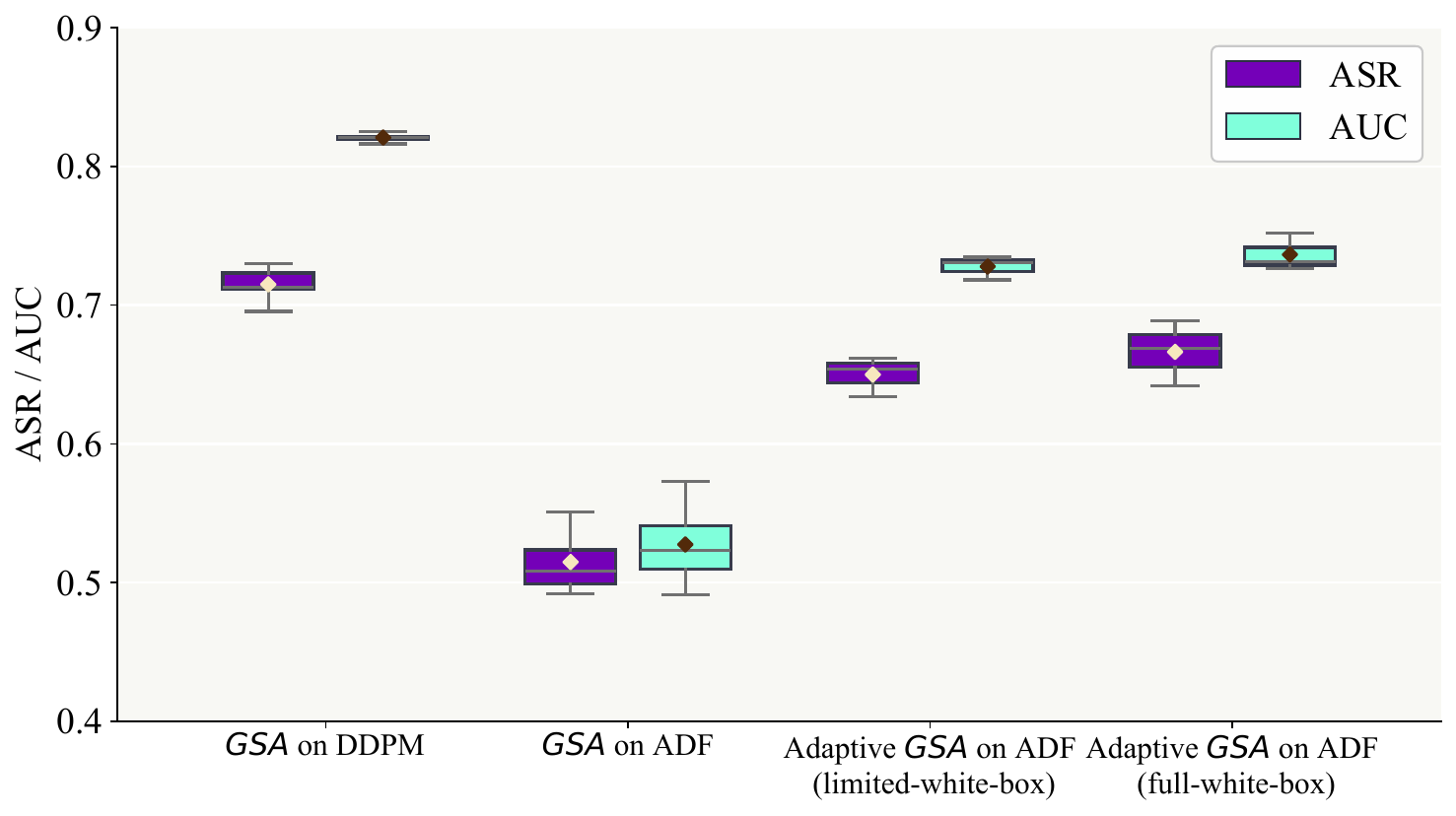}
    \caption{Evaluation of ADF against Adaptive Attacks.}
    \label{fig:adaptive_gsa}
\end{figure}

\textbf{Robustness under Adaptive Attack.}
To consider a stronger attack setting, where the adversary explicitly exploits knowledge of the freezing mask and places greater emphasis on unfrozen timesteps, we construct an adaptive attack based on GSA~\cite{pang2023white}. Specifically, gradient features are aggregated over the unfrozen timesteps for the subset assigned to the query sample. We consider two levels of attacker knowledge: (1) limited-white-box, where the attacker has access to the freezing mask and subset-partitioning rule; and (2) full-white-box, where the attacker is additionally given the training partition seed and can reproduce the exact subset assignment of the target-training samples. The adaptive attack adopts the standard shadow-model-based procedure to train an MIA classifier using the extracted mask-aware gradient features. We compare $GSA$ and adaptive $GSA$ on DDPM and ADF, as shown in Figure~\ref{fig:adaptive_gsa}: although the attack performance increases when $GSA$ is adapted to exploit the freezing mask in both limited-white-box and full-white-box settings, the ASR and AUC remain consistently lower than those of $GSA$ on the undefended DDPM. These results demonstrate that ADF provides effective privacy protection not only against standard $GSA$, but also against stronger method-aware adaptive attacks.

\textbf{Sensitivity Analysis across Partition Seeds.}
We further evaluate the sensitivity of ADF to the random seed used for data partitioning. As shown in Table~\ref{tab:partition_seed} in appendix, ADF exhibits stable privacy performance across randomly different partitions. Under three attacks, the standard deviations of ASR/AUC are 0.002/0.002, 0.005/0.006, and 0.007/0.010, respectively. More importantly, even the worst results across seeds remain below those of the undefended DDPM, related to Table~\ref{tab:privacy_protection_performance}. Similar reductions are observed for the TPR under low-FPR metrics. Meanwhile, generation quality is stable, with an average FID of $17.169\pm0.165$, which is consistently better than the DDPM FID of 18.227. These results indicate that ADF is not sensitive to a particular data partition and consistently maintains its privacy and utility performance across different partition seeds.

\subsubsection{Mechanism Analysis}
\begin{figure}[h]
    \centering
    \begin{subfigure}[h]{0.48\linewidth}
        \centering
        \includegraphics[width=\linewidth]{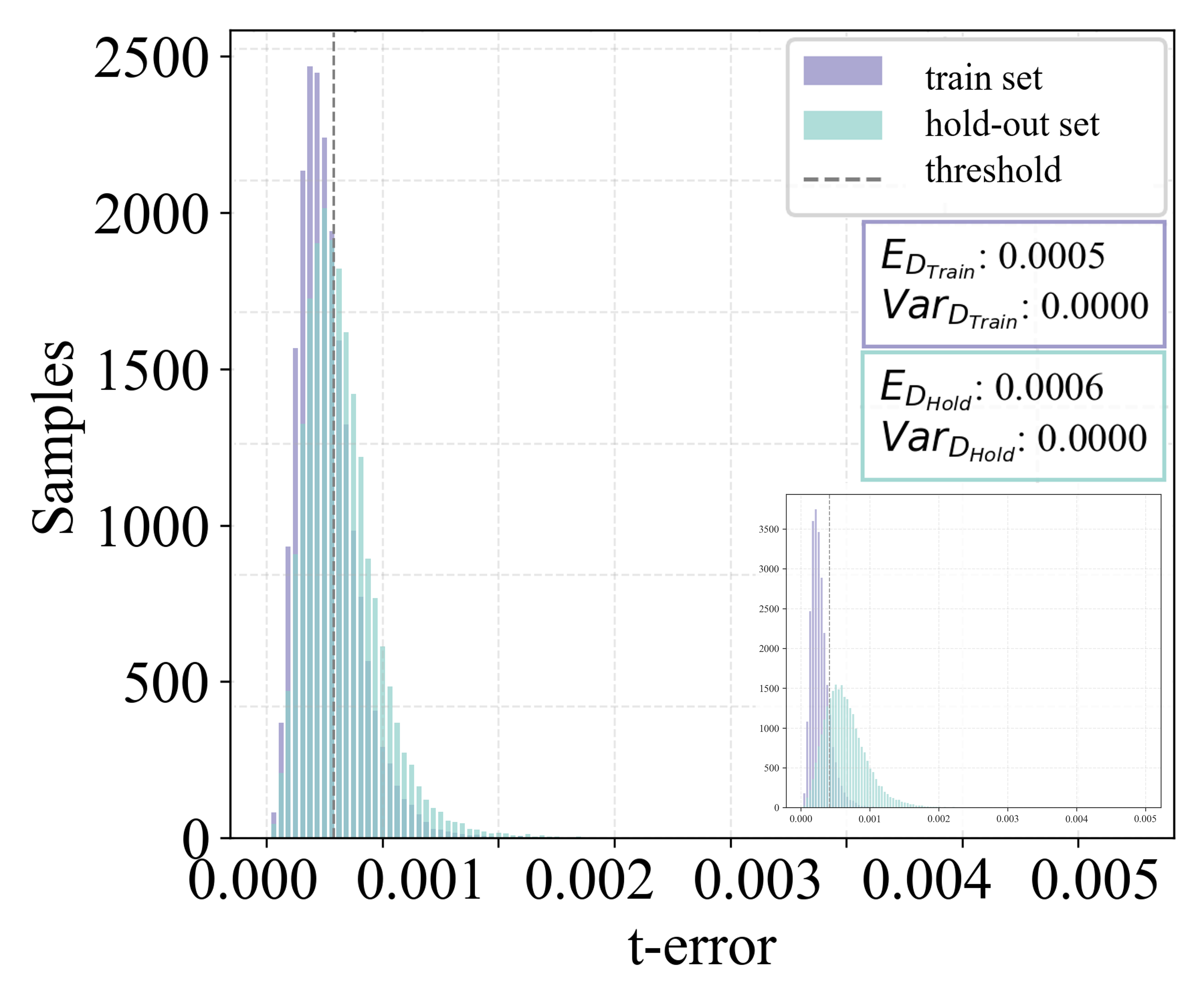}
        \caption{$SecMI_{stat}$}
        \label{fig:attack_feature_distribution_secmi_stat}
    \end{subfigure}
    \hfill
    \begin{subfigure}[h]{0.48\linewidth}
        \centering
        \includegraphics[width=\linewidth]{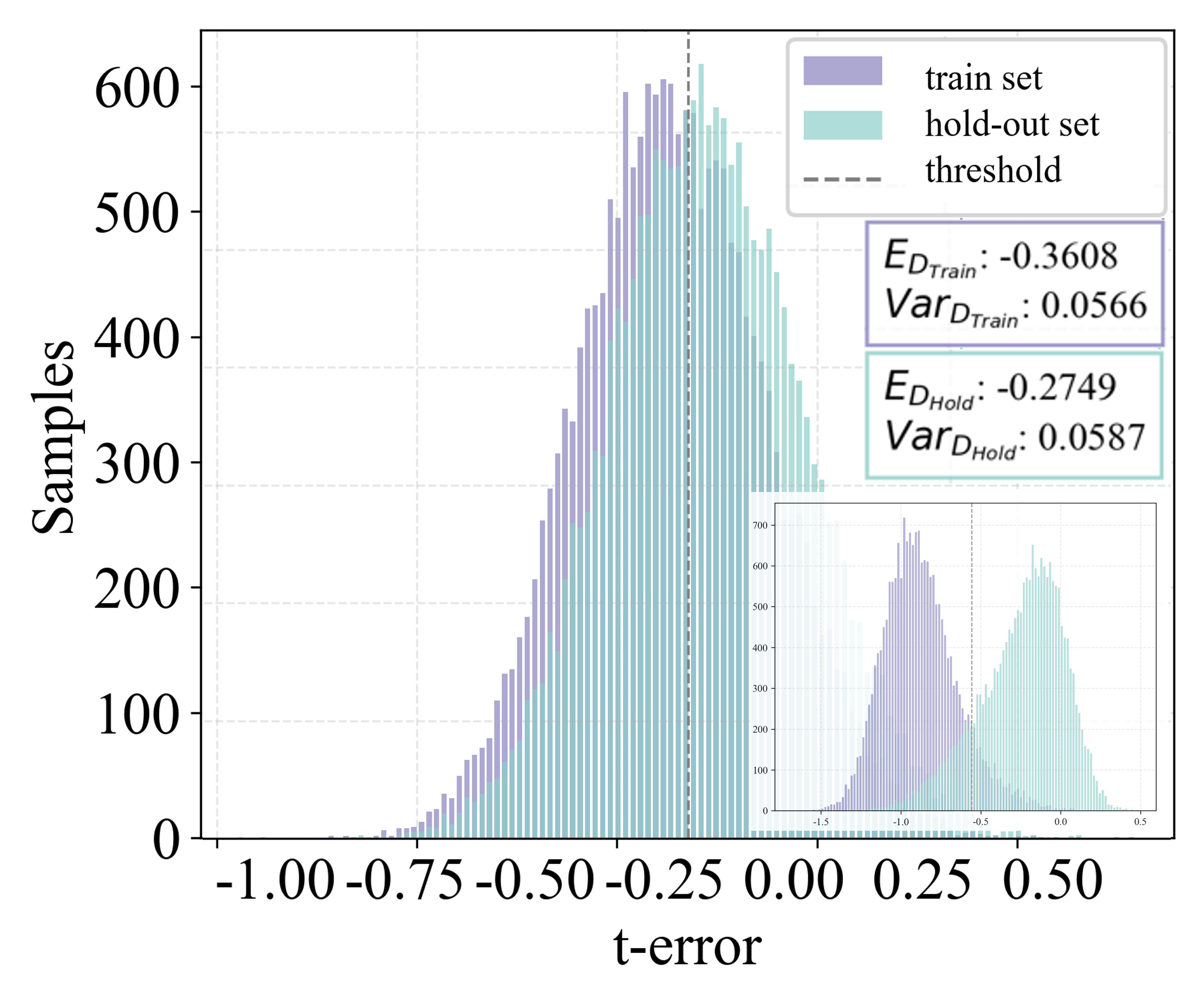}
        \caption{$SecMI_{nn}$}
        \label{fig:attack_feature_distribution_secmi_nn}
    \end{subfigure}
    \caption{Comparison of the distribution of attack feature: t-error.}
    \label{fig:attack_feature_distribution}
\end{figure}

We further analyze the defense mechanism from the perspective of attack characteristic distributions. As shown in Figure~\ref{fig:attack_feature_distribution}, before defending, the distribution of t-error, specifically, the estimated error of noise, exhibits a obvious deviation between member and nonmember samples, which enables attackers to exploit this statistical difference to infer membership. However, after applying ADF to DDPM, the distributions are almost completely overlapped, with similar means and variances. This indicates that ADF effectively eliminates the distinguishability of the model behavior between member and nonmember samples, thereby fundamentally undermining the basis for MIA based on statistical differences in attack features.

We also verify the connection between the loss reduction proxy and the final membership leakage, by measuring the loss reduction of each training sample and relating it to its final MIA vulnerability.
As shown in Table~\ref{tab:loss_reduction_risk}, members successfully identified by MIA exhibit substantially larger loss reduction than missed members with Weltch's $t=32.81$, $p<0.001$. Consistently, loss reduction is positively associated with binary MIA success ($r_{\mathrm{pb}}=0.231$) and continuous membership confidence (Pearson $r=0.207$, Spearman $\rho=0.112$), which are statistically significant. These results show that members exhibiting higher loss reduction during training tend to incur higher subsequent membership leakage, providing direct empirical support for using loss reduction as a informative risk-ranking proxy in risk-aware freezing policy of ADF.

\begin{table}[t]
    \centering
    \caption{Validation of loss reduction as a membership-risk proxy.}
    \label{tab:loss_reduction_risk}
    \resizebox{\linewidth}{!}{
        \begin{tabular}{lcc}
            \toprule
            \textbf{Items} & \textbf{Analysis} & \textbf{Value} \\
            \midrule
            
            \multirow{3}{*}{Loss reduction}
            & MIA-success members
            & 0.9722 \\
            
            & MIA-missed members
            & 0.9655 \\
            
            & Success vs.\ missed
            & 0.0067 ($t=32.806^{***}$) \\
            
            \midrule
            
            MIA Success 
            & Point biserial correlation
            & $r_{\mathrm{pb}}=0.231^{***}$ \\
            
            
            \multirow{2}{*}{Membership Confidence}
            & Pearson correlation
            & $r=0.207^{***}$ \\
            
            & Spearman correlation
            & $\rho=0.112^{***}$ \\
            
            \bottomrule
        \end{tabular}
    }
    \vspace{1mm}
    \begin{minipage}{\linewidth}
    \footnotesize
    \textit{Note.} All samples are true training members. ``MIA-success/missed members'' therefore refers to true members predicted as members and nonmembers, respectively. *$p<0.05$,**$p<0.01$,and ***$p<0.001$.
    \end{minipage}
\end{table}

\begin{figure}[h]
    \centering
    \begin{subfigure}[b]{\linewidth}
        \centering
        \includegraphics[width=0.96\linewidth]{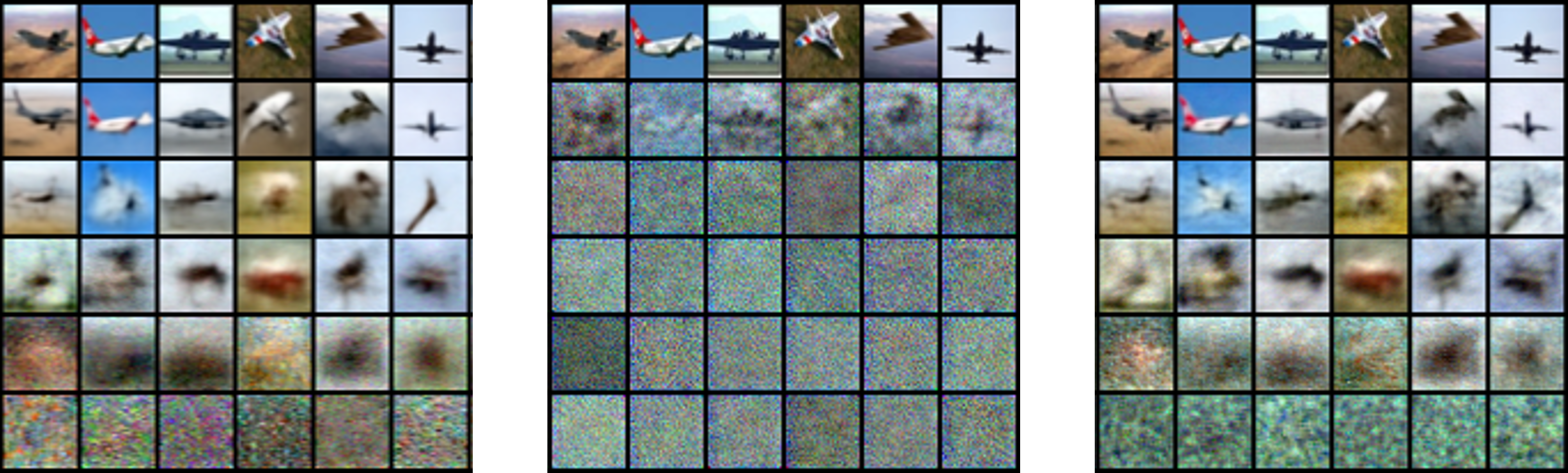}
        \caption{member samples}
        \label{fig:case_visualization_member}
    \end{subfigure}
    \begin{subfigure}[b]{\linewidth}
        \centering
        \includegraphics[width=0.96\linewidth]{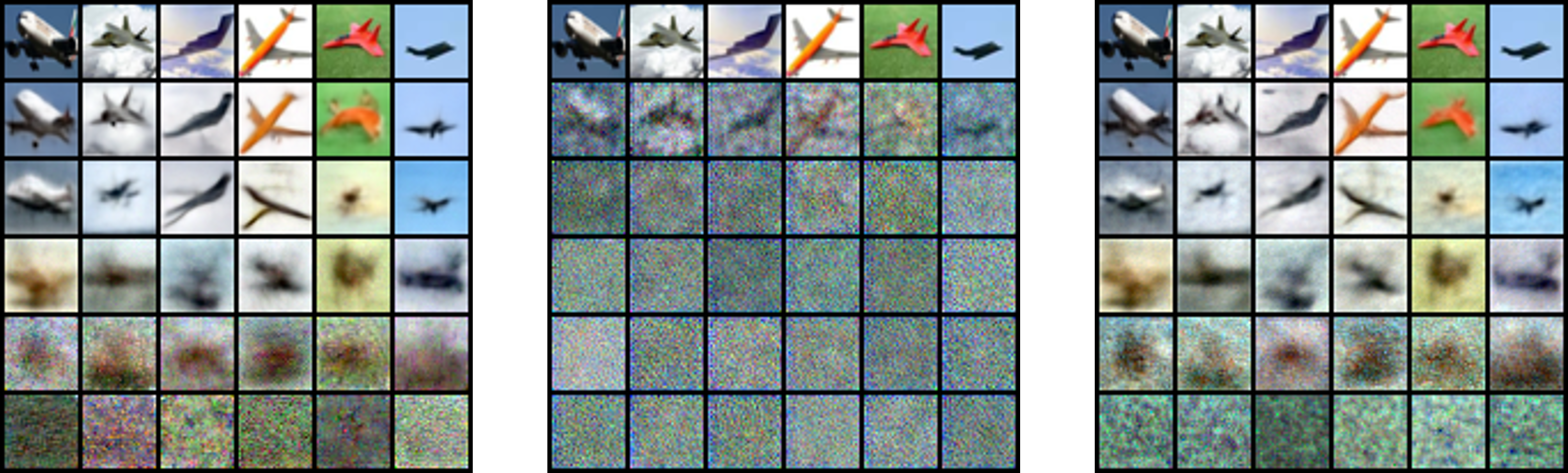}
        \caption{nonmember samples}
        \label{fig:case_visualization_nonmember}
    \end{subfigure}
    \caption{Comparison of the synthetic images: W/O (left), DP-SGD (middle), Ours (right). }
    \label{fig:case_visualization}
\end{figure}

Moreover, we examine the synthetic images and the intermediate denoising trajectories of ADF and baselines. We randomly select samples from both the training set and the hold-out set, apply the forward diffusion process to obtain their noisy versions, and record the intermediate images during the reverse denoising process of three models, i.e., W/O, DP-SGD, and Our ADF. Figure~\ref{fig:case_visualization_member} and Figure~\ref{fig:case_visualization_nonmember} illustrate the results of member samples and nonmember samples, respectively. In each plot, the first row shows the original clean images, the last row shows the corresponding pure random noise, and the rows in between show intermediate denoising results. 

As illustrated in Figure~\ref{fig:case_visualization}, ADF produces denoising trajectories that are visually closer to those of the undefended model. Compared with the undefended model, DP-SGD significantly disrupts the denoising trajectory. DP-SGD keeps injecting noise in training process, hence the intermediate denoised images become much blurrier and less semantically consistent, suggesting that the generative capability of the diffusion model is impaired. In contrast, the intermediate images generated by ADF gradually recover object-level structures and preserve more recognizable semantic details, while avoiding the excessive degradation observed under DP-SGD. The results indicate that ADF does not achieve privacy protection by globally weakening the diffusion model. Instead, it adaptively suppresses high-risk sample-timestep contributions, thereby reducing membership-specific behavioral differences while better preserving the original denoising dynamics and generation quality.

\subsubsection{Computational Efficiency}
We further evaluate the computational efficiency of different defense methods. Specifically, we report the number of trainable and extra parameters, total and relative training time, which reflect the additional structural overhead and runtime cost introduced by each defense. For a fair comparison, each target model is trained for 10,000 steps under the same setting. For knowledge distillation, the teacher model is also trained for 10,000 steps and is used to generate 2,000 samples to apply privacy distillation. For ADF, we apply the basic single-network setting with $G=1$.
Table~\ref{tab:computational_efficiency} in appendix compares the efficiency of different defenses. Overall, ADF achieves privacy protection without introducing extra trainable parameters in the main denoising training stage, as the mask matrix is constructed during pretraining and remains fixed thereafter. This avoids adding optimization complexity to already costly diffusion training. Although the pretraining-based risk estimation increases the overall computational cost compared to DDPM, ADF remains faster than DP-SGD and KD. These results demonstrate that ADF is a practical defense method for diffusion models with a favorable privacy-utility trade-off.

\section{Conclusion}
This paper proposes ADF, a novel defense framework for diffusion models to mitigate membership leakage risk. Compared to existing methods, ADF achieves effective privacy protection while maintaining high computational efficiency and utility. Experimental results demonstrate a favorable trade-off between privacy, utility, and efficiency across various datasets and attack settings. We evaluate ADF on standard diffusion models, e.g., DDPM, latent diffusion models, and state-of-the-art MMDiT models, Stable Diffusion 3, verifying the effectiveness of our methods across mainstream diffusion architectures. In addition, further studies can explore the extended application of ADF-paradigm against various privacy attacks, e.g., property inference attacks.

\bibliographystyle{ACM-Reference-Format}
\balance
\bibliography{ref}

\appendix
\section{Appendix}
\subsection{Privacy-utility trade-off performance of ADF across MIA settings.}

\begin{table*}[t]
\centering
\caption{Privacy-utility trade-off performance of ADF across MIA settings.
@1\% and @0.1\% denote TPR@1\%FPR and TPR@0.1\%FPR, respectively.}
\label{tab:privacy_protection_performance}
\setlength{\tabcolsep}{3.5pt}
\begin{tabular}{ccccccccccccccc}
\toprule
\multirow{2}{*}{\textbf{Dataset}} & \multirow{2}{*}{\textbf{Defense}} 
& \multirow{2}{*}{\textbf{FID}}
& \multicolumn{4}{c}{\textbf{$SecMI_{stat}$}} 
& \multicolumn{4}{c}{\textbf{$SecMI_{nn}$}}   
& \multicolumn{4}{c}{\textbf{$GSA$}} \\ 
\cmidrule(lr){4-7} \cmidrule(lr){8-11} \cmidrule(lr){12-15}
& & 
& \textbf{ASR} & \textbf{AUC} & \textbf{@1\%} & \textbf{@0.1\%}
& \textbf{ASR} & \textbf{AUC} & \textbf{@1\%} & \textbf{@0.1\%}
& \textbf{ASR} & \textbf{AUC} & \textbf{@1\%} & \textbf{@0.1\%} \\
\midrule

\multirow{6}{*}{CIFAR-10}
& W/O     & 18.227 & 0.678 & 0.737 & 0.033 & 0.004 & 0.720 & 0.789 & 0.082 & 0.013 & 0.772 & 0.858 & 0.239 & 0.084 \\
& Cutout  & 29.108 & 0.625 & 0.665 & 0.021 & 0.003 & 0.629 & 0.675 & 0.035 & 0.005 & 0.599 & 0.635 & 0.021 & 0.003 \\
& L2      & 27.201 & \textbf{0.518} & \textbf{0.521} & \textbf{0.009} & \textbf{0.001} & \textbf{0.505} & \textbf{0.504} & \textbf{0.011} & \textbf{0.001} & 0.543 & 0.559 & 0.014 & \textbf{0.001} \\
& DP-SGD  & 22.121 & 0.548 & 0.563 & 0.013 & 0.002 & 0.528 & 0.538 & 0.012 & \textbf{0.001} & 0.680 & 0.745 & 0.091 & 0.029 \\
& KD      & 55.203 & 0.523 & 0.529 & 0.010 & \textbf{0.001} & 0.507 & 0.505 & \textbf{0.011} & \textbf{0.001} & \textbf{0.533} & \textbf{0.548} & \textbf{0.013} & \textbf{0.001} \\
& Ours    & \textbf{17.672} & 0.540 & 0.553 & 0.012 & \underline{0.002} & 0.519 & 0.524 & \underline{0.012} & \textbf{0.001} & 0.587 & 0.622 & 0.034 & 0.005 \\
\midrule

\multirow{6}{*}{STL10\_U}
& W/O     & 20.551 & 0.621 & 0.664 & 0.028 & 0.002 & 0.675 & 0.737 & 0.069 & 0.010 & 0.740 & 0.821 & 0.145 & 0.030 \\
& Cutout  & 20.836 & 0.611 & 0.651 & 0.028 & 0.003 & 0.649 & 0.702 & 0.048 & 0.007 & 0.686 & 0.753 & 0.073 & 0.010 \\
& L2      & 31.226 & \textbf{0.510} & \textbf{0.512} & \textbf{0.012} & \textbf{0.001} & \textbf{0.502} & \textbf{0.499} & \textbf{0.010} & \textbf{0.001} & \textbf{0.521} & \textbf{0.532} & \textbf{0.015} & 0.002 \\
& DP-SGD  & 23.291 & 0.533 & 0.546 & 0.014 & \textbf{0.001} & 0.513 & 0.517 & 0.011 & \textbf{0.001} & 0.618 & 0.666 & 0.042 & 0.005 \\
& KD      & 49.812 & 0.513 & 0.518 & \textbf{0.012} & \textbf{0.001} & 0.503 & 0.502 & 0.011 & \textbf{0.001} & 0.539 & 0.555 & \textbf{0.015} & \textbf{0.001} \\
& Ours    & \textbf{20.052} & 0.555 & 0.577 & 0.016 & \textbf{0.001} & 0.539 & 0.554 & 0.016 & \underline{0.002} & 0.596 & 0.636 & \underline{0.033} & 0.005 \\
\midrule

\multirow{6}{*}{CelebA}
& W/O     & 46.564 & 0.547 & 0.565 & 0.018 & \textbf{0.000} & 0.545 & 0.563 & 0.016 & 0.002 & 0.592 & 0.627 & 0.028 & 0.003 \\
& Cutout  & 47.793 & 0.526 & 0.535 & \textbf{0.007} & 0.002 & 0.514 & 0.518 & 0.011 & \textbf{0.001} & 0.546 & 0.566 & 0.019 & 0.002 \\
& L2      & 54.892 & \textbf{0.502} & \textbf{0.499} & 0.009 & 0.001 & \textbf{0.501} & \textbf{0.498} & 0.011 & \textbf{0.001} & \textbf{0.497} & \textbf{0.498} & \textbf{0.010} & \textbf{0.001} \\
& DP-SGD  & 48.035 & 0.510 & 0.512 & 0.009 & 0.001 & 0.503 & 0.501 & 0.011 & \textbf{0.001} & 0.556 & 0.581 & 0.022 & 0.003 \\
& KD      & 51.045 & 0.506 & 0.507 & 0.009 & 0.001 & 0.502 & 0.500 & \textbf{0.010} & \textbf{0.001} & 0.509 & 0.512 & 0.011 & \textbf{0.001} \\
& Ours    & \textbf{46.353} & 0.507 & 0.510 & 0.011 & \underline{0.001} & 0.503 & 0.502 & \textbf{0.010} & \textbf{0.001} & 0.517 & 0.524 & \underline{0.011} & \textbf{0.001} \\
\midrule

\multirow{6}{*}{NWPU-RESISC45}
& W/O     & 23.484 & 0.597 & 0.620 & 0.010 & \textbf{0.000} & 0.614 & 0.656 & 0.042 & 0.005 & 0.648 & 0.709 & 0.061 & 0.007 \\
& Cutout  & 23.985 & 0.597 & 0.623 & 0.010 & \textbf{0.000} & 0.618 & 0.662 & 0.039 & 0.006 & 0.553 & 0.576 & 0.017 & 0.002 \\
& L2      & 38.774 & 0.534 & 0.530 & 0.009 & \textbf{0.000} & 0.543 & 0.540 & \textbf{0.009} & \textbf{0.001} & 0.544 & 0.540 & \textbf{0.010} & \textbf{0.001} \\
& DP-SGD  & 25.602 & 0.525 & 0.529 & 0.009 & \textbf{0.000} & 0.512 & 0.514 & 0.011 & \textbf{0.001} & 0.618 & 0.665 & 0.040 & 0.004 \\
& KD      & 50.624 & 0.515 & 0.518 & \textbf{0.008} & \textbf{0.000} & 0.507 & 0.506 & 0.011 & \textbf{0.001} & \textbf{0.523} & \textbf{0.529} & 0.011 & 0.002 \\
& Ours    & \textbf{22.232} & \textbf{0.507} & \textbf{0.505} & 0.010 & \textbf{0.000} & \textbf{0.503} & \textbf{0.497} & \underline{0.010} & \textbf{0.001} & \underline{0.527} & \underline{0.534} & 0.013 & \textbf{0.001} \\
\bottomrule
\end{tabular}
\end{table*}

\subsection{Sensitivity analysis of ADF across different partition seeds.}

\begin{table*}[t]
\centering
\caption{Sensitivity analysis of ADF across different partition seeds.}
\label{tab:partition_seed}
\resizebox{\textwidth}{!}{
\begin{tabular}{lcccccccc}
\toprule
& \multicolumn{2}{c}{\textbf{ASR}}
& \multicolumn{2}{c}{\textbf{AUC}}
& \multicolumn{2}{c}{\textbf{TPR@1\%FPR}}
& \multicolumn{2}{c}{\textbf{TPR@0.1\%FPR}} \\
\cmidrule(lr){2-3}
\cmidrule(lr){4-5}
\cmidrule(lr){6-7}
\cmidrule(lr){8-9}
\textbf{Attack}
& \textbf{Mean $\pm$ Std} & \textbf{Range}
& \textbf{Mean $\pm$ Std} & \textbf{Range}
& \textbf{Mean $\pm$ Std} & \textbf{Range}
& \textbf{Mean $\pm$ Std} & \textbf{Range} \\
\midrule

$SecMI_{stat}$
& $0.585 \pm 0.002$ & $[0.582,\,0.587]$
& $0.614 \pm 0.002$ & $[0.611,\,0.617]$
& $0.0160 \pm 0.0010$ & $[0.0150,\,0.0172]$
& $0.0017 \pm 0.0003$ & $[0.0012,\,0.0022]$ \\

$SecMI_{nn}$
& $0.572 \pm 0.005$ & $[0.566,\,0.578]$
& $0.600 \pm 0.006$ & $[0.595,\,0.608]$
& $0.0215 \pm 0.0007$ & $[0.0206,\,0.0225]$
& $0.0027 \pm 0.0005$ & $[0.0021,\,0.0034]$ \\

$GSA$
& $0.587 \pm 0.007$ & $[0.579,\,0.596]$
& $0.636 \pm 0.010$ & $[0.622,\,0.644]$
& $0.0313 \pm 0.0040$ & $[0.0276,\,0.0374]$
& $0.0055 \pm 0.0011$ & $[0.0042,\,0.0069]$ \\

\midrule
\multicolumn{3}{l}{\textbf{FID}}
& \multicolumn{3}{c}{Mean $\pm$ std: $17.169 \pm 0.165$}
& \multicolumn{3}{c}{Range: $[16.961,\,17.396]$} \\

\bottomrule
\end{tabular}
}
\end{table*}

\subsection{Computational efficiency comparison of different defense methods.}

\begin{table}[h]
\centering
\small
\setlength{\tabcolsep}{3pt}
\caption{Computational efficiency comparison of different defense methods.}
\label{tab:computational_efficiency}
\resizebox{\linewidth}{!}{
\begin{tabular}{lcccc}
\toprule
\textbf{Method} 
& \textbf{Trainable} 
& \textbf{Extra} 
& \textbf{Time} 
& \textbf{Runtime} \\
& \textbf{Params} 
& \textbf{Params} 
& \textbf{(s)} 
& \textbf{vs. DDPM} \\
\midrule
W/O     & 35.75M & 0      & 1684.36 & 1.00$\times$ \\
Cutout  & 35.75M & 0      & 1695.37 & 1.01$\times$ \\
L2      & 35.75M & 0      & 1707.31 & 1.01$\times$ \\
DP-SGD  & 35.75M & 0      & 4278.79 & 2.54$\times$ \\
KD      & 71.49M & 61.31M & 7458.01 & 4.43$\times$ \\
Ours    & 35.75M & 0      & 4064.40 & 2.41$\times$ \\
\bottomrule
\end{tabular}
}
\end{table}

\section{Open Science}

\textbf{Artifacts.}
The artifacts required to evaluate the core contributions of this paper include:
(i) the implementation of our proposed defense framework,
(ii) scripts for training defended and undefended diffusion models,
(iii) scripts for running membership inference attacks, including $SecMI_{stat}$, $SecMI_{nn}$, and $GSA$, 
and (iv) scripts for computing generative utility metrics such as FID.

\textbf{Artifacts Access.}
Our code is available at \url{https://github.com/JialuRita/AdaptiveDiffusionFreezing-ADF}. It includes a README file with step-by-step instructions for environment setup, dataset preparation, model training, utility evaluation, membership inference attacks.

\textbf{Datasets.}
Our experiments use public image datasets, including CIFAR-10, CelebA, STL10\_U, and NWPU-RESISC45. We do not redistribute the original datasets due to the file size and copyright. Instead, we provide scripts to download and pre-process the datasets for our experiment settings.

\textbf{Models and Checkpoints.}
Due to file size limitations, we do not include pre-trained model checkpoints in the repository. Instead, we provide the training and evaluation scripts, command-line instructions, and execution guidelines required to reproduce the reported results.

\section{Ethical Considerations}
This work studies membership inference attacks (MIAs) against diffusion models with the goal of improving the privacy protection. Although MIAs can potentially be misused to infer whether a sensitive sample was included in a model's training data, our study focuses on defensive purposes. Specifically, we use existing attack methods only as evaluation tools to quantify membership leakage risks and to assess the effectiveness of the proposed defense. We do not facilitate attacks against real-world models and systems, nor do we release any private or sensitive training data.

All experiments are conducted on publicly available benchmark datasets that are widely used in generative modeling research. The datasets are used solely for controlled empirical evaluation, and we do not attempt to identify real individuals or infer sensitive attributes from generated samples. 

\end{document}